\documentclass[12pt,draftclsnofoot,onecolumn]{IEEEtran}
\usepackage{cite}
\usepackage{amsmath,amssymb,amsfonts}
\usepackage{algorithmic}
\usepackage[ruled,linesnumbered]{algorithm2e}
\usepackage{scalerel}
\usepackage{graphicx}
\usepackage{textcomp}
\usepackage{xcolor}
\usepackage{amsbsy}
\usepackage{gensymb}
\usepackage{enumitem}
\usepackage{amsthm}
\usepackage{amsmath,bm}
\usepackage{caption}
\usepackage{subcaption}
\usepackage{soul}
\usepackage[multiple]{footmisc} 

\makeatletter
\renewcommand{\@IEEEsectpunct}{~}
\makeatother

\theoremstyle{remark} 
\newtheorem{remark}{Remark}

\def\BibTeX{{\rm B\kern-.05em{\sc i\kern-.025em b}\kern-.08em
    T\kern-.1667em\lower.7ex\hbox{E}\kern-.125emX}}
\begin{document}
%
\title{Roadside IRS-Aided Vehicular Communication: Efficient Channel Estimation and Low-Complexity Beamforming Design}
\author{Zixuan~Huang, \IEEEmembership{Student Member,~IEEE,} Beixiong~Zheng, \IEEEmembership{Member,~IEEE,} and~Rui~Zhang, \IEEEmembership{Fellow,~IEEE}
\thanks{Z.~Huang is with the NUS Graduate School, National University of Singapore, Singapore 119077, and also with the Department of Electrical and Computer Engineering, National University of Singapore, Singapore 117583 (e-mail: huang.zixuan@u.nus.edu).}
\thanks{B.~Zheng is with the School of Microelectronics, South China University of Technology, Guangzhou 511442, China  (e-mail: bxzheng@scut.edu.cn). He was with the Department of Electrical and Computer Engineering, National University of Singapore, Singapore 117583.}
\thanks{R.~Zhang is with The Chinese University of Hong Kong, Shenzhen, and Shenzhen Research Institute of Big Data, Shenzhen, China 518172 (e-mail: rzhang@cuhk.edu.cn). He is also with the Department of Electrical and Computer Engineering, National University of Singapore, Singapore 117583 (e-mail: elezhang@nus.edu.sg).
}
}

%
%
%



\maketitle
\begin{abstract}
Intelligent reflecting surface (IRS) has emerged as a promising technique to control wireless propagation environment for enhancing the communication performance cost-effectively.
However, the rapidly time-varying channel in high-mobility communication scenarios such as vehicular communication renders it challenging to obtain the instantaneous channel state information (CSI) efficiently for IRS with a large number of reflecting elements.
In this paper, we propose a new roadside IRS-aided vehicular communication system to tackle this challenge.
Specifically, by exploiting the symmetrical deployment of IRSs with inter-laced equal intervals on both sides of the road and the cooperation among nearby IRS controllers, we propose a new two-stage channel estimation scheme with off-line and online training, respectively, to obtain the static/time-varying CSI required by the proposed low-complexity passive beamforming scheme efficiently.
The proposed IRS beamforming and online channel estimation designs leverage the existing uplink pilots in wireless networks and do not require any change of the existing transmission protocol. 
Moreover, they can be implemented by each of IRS controllers independently, without the need of any real-time feedback from the user's serving BS.
Simulation results show that the proposed designs can efficiently achieve the high IRS passive beamforming gain and thus significantly enhance the achievable communication throughput for high-speed vehicular communications.
\end{abstract}


\section{Introduction} 
Both academia and industry have devoted significant effort to achieving high-performance communications for high-speed vehicles in transportation infrastructures including vehicle-to-vehicle (V2V), vehicle-to-infrastructure (V2I), vehicle-to-pedestrian (V2P), and vehicle-to-network (V2N) communications (collectively termed as vehicle-to-everything (V2X) communications) \cite{cv2x0,cv2x01,cv2x2}.
However, the fast-growing demands for V2X communications (e.g., passenger infotainment, autonomous driving, intelligent transportation systems, and so on) may not be fully realized by today's fifth-generation (5G) wireless networks \cite{chall1}, and the rapidly time-varying wireless channel due to high-mobility users is still the ultimate bottleneck in achieving the high-capacity, ultra-reliable, and low-latency V2X communications. 
To overcome this barrier, various wireless techniques have been proposed such as dynamic resource allocation, active beamforming, diversity, adaptive modulation/coding, etc., to either adapt to the random and time-varying wireless channel or compensate for its fading effects \cite{urllc0,urllc1}.
Since these techniques are applied at wireless transceivers, they cannot fully mitigate the wireless channel impairments to guarantee the stringent quality-of-service (QoS) requirement of V2X communications.


Recently, intelligent reflecting surface (IRS) \cite{irs1,irs_tut} and its equivalents (such as reconfigurable intelligent surface (RIS) \cite{ris2}) have emerged as a cost-effective solution to achieve smart and configurable wireless propagation environment by dynamically tuning signal reflection and thereby enhance the wireless network performance. 
Specifically, IRS is a digitally controlled metasurface that is composed of many passive reflecting elements, each being able to independently adjust the amplitude and/or phase of the incident signal in real time \cite{hard1,hard2}. 
Thus, different from conventional wireless communication techniques employed at transceivers, IRS is able to flexibly reshape the wireless propagation channel for a variety of purposes, such as bypassing obstacles/obstructions \cite{irs1}, refining wireless channel realizations/distributions \cite{vehicle-side-irs}, improving the multi-antenna/multiuser channel rank condition \cite{irsfun2}, among others. 
Moreover, IRS dispenses with radio frequency (RF) chains and operates in full-duplex mode with passive reflection only, which thus features low hardware cost and power consumption, and is deemed a promising technology for the next-/sixth-generation (6G) wireless networks.
As such, IRS has spurred intense research interest and been thoroughly investigated for various wireless systems, such as multiple-input multiple-output (MIMO) communications \cite{mimo1,mimo2}, orthogonal frequency division multiplexing (OFDM) based systems \cite{ofdm1,ofdm2,ofdm3}, non-orthogonal multiple access (NOMA) \cite{noma1,noma2}, simultaneous wireless information and power transfer (SWIPT) \cite{swipt1,swipt2}, mobile edge computing \cite{mec1,mec2}, etc.


To achieve effective control over the wireless propagation environment by IRSs, the acquisition of accurate channel state information (CSI) in IRS-aided wireless communication systems is crucial, which, however, is practically challenging to realize due to the lack of signal processing capabilities at IRS reflecting elements as well as their massive number in practice.
Although the IRS\textrightarrow base station (BS)/user channels cannot be separately estimated by IRSs that are fully passive, the cascaded user\textrightarrow IRS\textrightarrow BS channels can be estimated at the BS based on the pilot symbols sent by the users with properly designed IRS reflection patterns over time \cite{ofdm2}.
However, the acquisition of such CSI in IRS-aided systems may require a prohibitively high training overhead that is in general proportional to the number of reflecting elements and thus can severely degrade the data communication throughput.
To reduce the channel training overhead, various methods have been developed in the literature for IRS, such as IRS elements grouping \cite{ofdm1,grouping1}, reference-user-based channel estimation \cite{liu,bx_add}, anchor-aided channel estimation \cite{guan}, channel-sparsity-based estimation \cite{sparse1}, and so on (see e.g., \cite{irs_tut,est_2} and the references therein).

However, existing works on IRS have mostly focused on assisting the communications of low-mobility users with slow fading channels to/from the IRS, which may not be applicable to high-mobility scenarios such as high-speed vehicular communication. In this case, due to the vehicle's high speed and environment's random scattering, the transmitted signal from the user usually arrives at the BS over multiple propagation paths subjected to independently and rapidly time-varying phase shifts due to different Doppler frequencies, thus resulting in a superimposed fast fading channel (i.e., both the amplitude and phase shift of the overall channel vary substantially over time).
As a result, the reliability and throughput of data communication between the vehicle and its serving BS can be severely degraded. 
In \cite{vehicle-side-irs}, the authors considered the high-mobility communication aided by the vehicle-side IRS and proposed a low-complexity channel estimation scheme to track the BS\textrightarrow IRS channel variation efficiently. 
However, the coverage of a vehicle-side IRS is limited to the users inside the vehicle only, which thus needs to be separately employed in each vehicle and may incur a high cost to the vehicle manufacturing.
In \cite{bass,dsce,kf1,kf2}, the authors considered an alternative approach, in which IRSs are deployed on the roadside at regular intervals to assist in the high-speed vehicular communication in a consecutive manner. 
In \cite{sb1,dll,sb2}, the authors considered the two-timescale IRS channel estimation/beamforming design to reduce the training/signaling overhead by exploiting the static BS\textrightarrow IRS channel.
However, these studies require additional pilot symbols from the users dedicated for estimating the IRS channels, which can result in substantial training overhead and also needs to modify the existing transmission protocol of mobile users (e.g., that of the massive MIMO communication in the cellular uplink \cite{cell_net}) which only estimates the user\textrightarrow BS channel without IRS.
Moreover, in the existing works, the IRS reflection is usually designed based on the CSI acquired at the BS, which needs to be fed back to each IRS controller to adjust its corresponding IRS's reflection, thus inevitably incurring feedback delay. Such delay may render the designed IRS reflection outdated due to the rapid channel variation and thus less effective in high-mobility scenarios.

\begin{figure}
\centering
\includegraphics[width=0.8\textwidth]{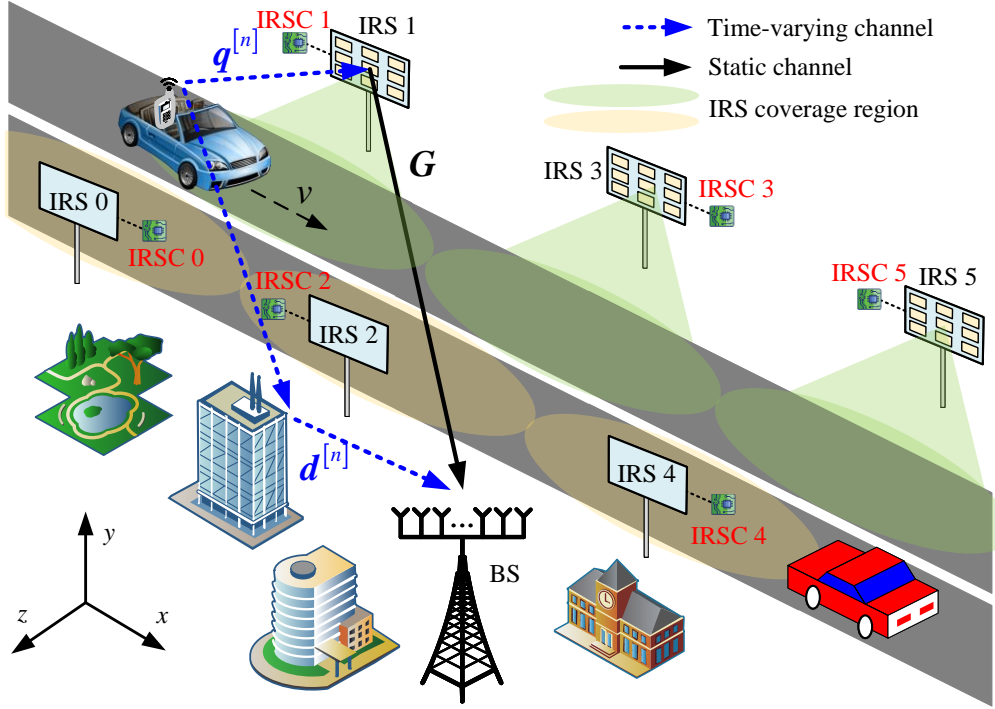}
\caption{Roadside IRS-aided communication for serving high-speed vehicles.}
\label{config}
\end{figure}

Motivated by the above, we propose in this paper a new roadside IRS-aided vehicular communication system, as shown in Fig.~\ref{config}, to achieve efficient channel estimation and low-complexity passive beamforming for IRS, which require neither modification of the existing uplink transmission protocol nor real-time feedback from the BS to each serving IRS, thus making the roadside IRS-aided system practically appealing for enhancing high-speed vehicular communication performance efficiently.
The main contributions of this paper are summarized as follows.
\begin{itemize}
    \item 
    First, we propose a low-complexity passive beamforming design for each serving IRS to maximize the power of the IRS-reflected channel from the user to the BS in the uplink. Specifically, we first design an initial passive beamforming vector for each serving IRS based on its off-line estimated CSI without any user involvement. 
    Then, based on the CSI of the user\textrightarrow IRS channel acquired in real time, the actual passive beamforming vector is designed accordingly for assisting the user's uplink communication with the BS.
    \item
    Next, we propose a new two-stage channel estimation scheme to obtain the CSI required by the proposed passive beamforming design at the serving IRS. In the first off-line estimation stage, the BS estimates the serving IRS\textrightarrow BS channel and serving IRS\textrightarrow its controller\footnote{IRS controller is attached to each IRS for controlling its signal reflection as well as communicating with the associated BS/user for exchanging control signals in IRS-aided communications. Thus, the IRS controller needs to possess both transmit and receive RF chains and it can also send/receive pilot signals for helping channel estimation as considered in this paper.}\footnote{Conventional active relay also possesses RF chains. However, the active multi-antenna relay generally requires more RF chains than the IRS/RIS controller (with one RF chain only), which leads to higher hardware cost and power consumption. Moreover, the active relaying system practically operates in the half-duplex mode to save hardware cost, which, however, suffers from low spectral efficiency. On the other hand, the full-duplex relay requires much higher hardware cost and sophisticated signal processing for self-interference cancellation. In contrast, our proposed IRS/RIS controller-assisted system operates in full-duplex mode and is free of antenna noise amplification as well as self-interference. In addition, active relays forward the information signals while IRS/RIS controller transmits/receives signals for controlling/beam design purposes only.} channel based on the training signals sent by the controllers of two nearby IRSs symmetrically deployed at the opposite roadside of the serving IRS, and feeds back the initially designed IRS passive beamforming and the latter estimated CSI to the serving IRS's controller.
    In the second online estimation stage, the serving IRS's controller first estimates the instantaneous CSI of the user\textrightarrow IRS channel based on the off-line estimated IRS\textrightarrow controller CSI as well as the received uplink pilot signals from the user before the IRS starts to serve the moving user, without the need of changing the existing uplink transmission protocol.
    Then, when the user enters into the coverage region of the serving IRS, a novel passive beam prediction algorithm is proposed for the serving IRS's controller to predict the time-varying user\textrightarrow IRS CSI based on the previously estimated CSI for updating the IRS passive beamforming over time. Note that the above process is implemented at each serving IRS independently without the need of any real-time instruction/feedback from the BS.
    \item
    Last, we provide extensive simulation results to evaluate the performance of the proposed roadside IRS-aided vehicular communication system. 
    We show that our proposed design is effective in enhancing the IRS-aided channel power in the uplink communication significantly and thus yields a considerably higher achievable rate in the high-mobility scenario, as compared to the conventional design for IRS-aided low-mobility communications.
    The effects of various key system parameters on the achievable throughput, such as the channel training overhead, IRS coverage distance, number of IRS reflecting elements, and vehicle speed, are thoroughly investigated and discussed. 
\end{itemize}

The rest of this paper is organized as follows.
Section II presents the system model for the proposed roadside IRS-aided vehicular communication system. In Section III, we propose a low-complexity IRS passive beamforming design. In Section IV, we propose a new two-stage channel estimation scheme for enabling the proposed passive IRS beamforming design. Simulation results and discussions are presented in Section V. Finally, conclusions are drawn in Section VI.

{\emph{Notations:}} 
Upper-case and lower-case boldface letters denote matrices and column vectors, respectively. Upper-case calligraphic letters (e.g., $\mathcal{T}$) denote discrete and finite sets. Superscripts $(\cdot)^{T}$, $(\cdot)^{*}$, and $(\cdot)^\dagger$ stand for the transpose, conjugate, and Moore-Penrose inverse operations, respectively.
$\left\|\cdot\right\|$ denotes the $l_2$ norm.
$\left\|\cdot\right\|_2$ denotes the Frobenius norm.
$\lceil \cdot \rceil$ denotes the ceiling operation.
$\mathbb{R}^{x \times y}$ denotes the space of $x \times y$ real matrices.
$\mathbb{C}^{x \times y}$ denotes the space of $x \times y$ complex matrices.
$[\cdot]_{i,j}$ denotes the $(i,j)$-th element of a matrix.
$\operatorname{diag}(\boldsymbol{x})$ denotes a square diagonal matrix with the elements of $\boldsymbol{x}$ on the main diagonal.
$\operatorname{vec}(\boldsymbol{X})$ denotes the vectorization operation of a matrix $\boldsymbol{X}$.
$\otimes$ denotes the Kronecker product. 
$\odot$ denotes the Hadamard product. 
$\oslash$ denotes the element-wise division. 
$\boldsymbol{\mathrm{I}}_{x}$ denotes an identity matrix with its dimension of $x$.
$\boldsymbol{1}_{M}$ denotes an all-one vector with its dimension of $M$.
$\mathcal{O}(\cdot)$ describes the order of complexity.
The distribution of a circularly symmetric complex Gaussian (CSCG) random variable with zero mean and variance $\sigma^2$ is denoted by $\mathcal{N}_{c}\left(0, \sigma^{2}\right)$; and $\sim$ stands for “distributed as”. The main symbols used in this paper are listed in Table I with their corresponding meanings given.

\begin{table*}
\centering
\caption{List of Main Symbols and Their Physical Meanings}
\begin{tabular}{|c | c |} 
 \hline
 \textbf{Symbol} & \textbf{Physical meaning}  \\ [0.5ex] 
 \hline\hline
 IRS\textsubscript{$k$} & IRS $k$  \\ 
 \hline
 C\textsubscript{$k$} & Controller of IRS $k$  \\ 
 \hline
 $\boldsymbol{d}^{[n]}$ & User\textrightarrow BS channel in block $n$  \\ 
 \hline
 $\boldsymbol{q}^{[n]}$ & User\textrightarrow serving IRS channel in block $n$  \\
 
 $a^{[n]}$ & Path gain of $\boldsymbol{q}^{[n]}$  \\ 
 
 $\theta^{[n]}/\phi^{[n]}$ & Elevation/azimuth angle of arrival (AoA) from the user to the serving IRS in block $n$ \\ 
 
 $\vartheta^{[n]}/\psi^{[n]}$ & Array phase at the serving IRS along the $x/y$-axis in $\boldsymbol{q}^{[n]}$  \\

 \hline
 $\boldsymbol{G}$ & Serving IRS\textrightarrow BS channel  \\
 
 $L$ & Number of paths in $\boldsymbol{G}$  \\ 
 
 $a_l$ & Path gain of the $l$-th path in $\boldsymbol{G}$   \\ 
 
 $\theta_{\mathrm{B},l}$ & AoA of the $l$-th path at the BS in $\boldsymbol{G}$  \\ 
 
 $\zeta_{l}$ & Array phase of the $l$-th path at the BS in $\boldsymbol{G}$  \\ 
 
 $\theta_{l}/\phi_{l}$ & Elevation/azimuth angle of departure (AoD) of the $l$-th path at the serving IRS in $\boldsymbol{G}$  \\ 
 
 $\vartheta_{l}/\psi_{l}$ & Array phase of the $l$-th path at the serving IRS along the $x/y$-axis in $\boldsymbol{G}$  \\ 
 \hline
 
 $\boldsymbol{b}_{k}$ & C\textsubscript{$k$}\textrightarrow serving IRS channel  \\ 
 
 $\dot{a}_{k}$ & Path gain of $\boldsymbol{b}_{k}$  \\ 
 
 $\dot{\theta}_{k}/\dot{\phi}_{k}$ & Elevation/azimuth AoA from C\textsubscript{$k$} to the serving IRS  \\ 
 
 $\dot{\vartheta}_{k}/\dot{\psi}_{k}$ & Array phase at the serving IRS along the $x/y$-axis in $\boldsymbol{b}_{k}$  \\ 
 \hline
 
 $\boldsymbol{R}_k$ & Cascaded C\textsubscript{$k$}\textrightarrow serving IRS\textrightarrow BS channel  \\
 
 $a_{l,k}$ & Product path gain of the $l$-th path in $\boldsymbol{R}_k$   \\ 
 $\vartheta_{l,k}/\psi_{l,k}$ & Effective array phase of the $l$-th path at the serving IRS along the $x/y$-axis in $\boldsymbol{R}_k$  \\ 
 \hline
 
 $\boldsymbol{Q}^{[n]}$ & Cascaded user\textrightarrow serving IRS\textrightarrow BS channel in block $n$  \\
 \hline
 $\boldsymbol{h}^{[n]}$ & Overall user\textrightarrow BS channel in block $n$  \\
 \hline
 $\boldsymbol{\nu}^{[n]}$ & Reflection coefficient vector of the serving IRS in block $n$  \\ 
 \hline
 $\boldsymbol{\bar{\nu}}_\mathrm{ini}$ & Initial passive beamforming vector of the serving IRS  \\ 
 \hline
\end{tabular}
\end{table*}

\section{System Model} \label{sec_1}
\subsection{Roadside IRS-Aided Vehicular Communication}
As shown in Fig.~\ref{config}, we consider a high-mobility vehicular communication system aided by multiple roadside IRSs deployed on both sides of a road, on which vehicles travel bidirectionally on their corresponding lanes.
Without loss of generality, we consider the communication between one BS and a mobile user\footnote{If more than one users need to be served by each IRS at the same time, then the IRS can be split into multiple sub-surfaces each serving one user simultaneously, while the passive beamforming gain for each user will be reduced.} (a vehicle or any user inside the vehicle), which is assisted by the IRSs deployed on one side of the road (labeled with odd numbers as shown in Fig.~\ref{config})\footnote{Due to the half-space signal reflection of each IRS, we consider that this user is served by the BS located on the other side of the road (see Fig.~\ref{config}) so that the IRSs labeled with odd numbers can effectively reflect the signals between the BS and the user.}.
Assuming time-division duplexing (TDD) based communication between the user and its serving BS, we focus on the uplink communication from the user to the BS in this paper, while the results of this paper can be also applied to the downlink communication in the reverse direction by exploiting the uplink-downlink channel reciprocity.
For the purpose of exposition, we assume that the user is moving along the $x$-direction with a high speed of $v$.
We also assume that the BS is equipped with a uniform linear array (ULA) consisting of $N_{\mathrm{B}}$ antennas and the mobile user is equipped with a single antenna. 
Each IRS is equipped with a uniform planar array (UPA) composed of $M=M_x \times M_y$ reflecting elements placed in the $x-y$ plane in the three-dimensional (3D) Cartesian coordinate system as shown in Fig.~\ref{config}, which is connected to an IRS controller (IRSC) that is capable of adjusting the corresponding IRS elements' individual reflection amplitude and/or phase shift as well as processing/exchanging (control/channel) information with its assisted BS via a separate and reliable wireless link. 
In this paper, we consider block-fading channels for all the channels associated with the mobile user, which are assumed to remain approximately constant during each transmission block, but may vary from block to block due to the user's high mobility. The duration of each block is denoted as $T_b$, which is chosen to be sufficiently small as compared to the minimum coherence interval of all the channels involved.

\subsection{Channel Model} 
Without loss of generality, we focus on one transmission frame consisting of $N$ blocks (indexed by the set $\mathcal{N} \triangleq \{1,\ldots,N\}$) when the vehicle/user is passing by its serving IRS (assumed to be IRS 1) while communicating with the BS in the uplink.
For brevity and without loss of generality, we drop the index of IRS 1 in the following.
Accounting for the user's high mobility, we let $\boldsymbol{d}^{[n]} \in \mathbb{C}^{N_{\mathrm{B}} \times 1}$ and $\boldsymbol{q}^{\left[n\right]} \in \mathbb{C}^{M \times 1}$ denote the time-varying channels for the user\textrightarrow BS and user\textrightarrow IRS links in block $n$. 
In contrast, given the fixed locations of the IRS and BS, we let $\boldsymbol{G} \in \mathbb{C}^{N_{\mathrm{B}} \times M}$ denote the IRS\textrightarrow BS channel, which is assumed to remain static during the transmission frame of interest.

For convenience, let $\boldsymbol{e}\left( \phi,\bar{M} \right) = \left[1,e^{j \pi \phi}, \ldots, e^{j( \bar{M}-1 )\pi \phi}\right]^T$ denote the one-dimensional (1D) steering vector function, where $j = \sqrt{-1}$ denotes the imaginary unit, $\phi$ denotes the phase difference (normalized to $\pi$) between any two adjacent antennas/elements, and $\bar{M}$ denotes the number of antennas/elements of the 1D array of interest. Under the ULA model, the array response vector at the BS is denoted by $\boldsymbol{e}\left( \zeta,N_{\mathrm{B}} \right)$, where $\zeta = \frac{2 d_\mathrm{B}}{\lambda} \cos {\theta} \in \left[-\frac{2d_\mathrm{B}}{\lambda},\frac{2d_\mathrm{B}}{\lambda}\right]$ denotes the array phase with $\theta \in \left[0,\pi \right]$ being the angle of arrival (AoA), $d_\mathrm{B}$ denotes the spacing between any two adjacent antennas, and $\lambda$ denotes the signal wavelength.
Under the UPA model, the array response vector at the IRS is expressed as the Kronecker product of two 1D steering vector functions in the $x$- and $y$-axis directions, respectively, i.e.,
\begin{equation}\label{response}
\boldsymbol{u}\left( \vartheta,\psi \right) =
\boldsymbol{e}\left(  \vartheta ,M_x \right) \otimes \boldsymbol{e}\left( \psi ,M_y \right),
\end{equation} 
where $\theta \in \left[0,\pi\right]$ and $\phi \in \left[0,2\pi\right)$ respectively denote the elevation and azimuth AoAs or angles of departure (AoDs) at the IRS, $\vartheta = \frac{2d_\mathrm{I}}{\lambda} \cos \theta \cos \phi \in \left[-\frac{2d_\mathrm{I}}{\lambda},\frac{2d_\mathrm{I}}{\lambda}\right]$ and $\psi = \frac{2d_\mathrm{I} }{\lambda} \cos \theta \sin \phi \in \left[-\frac{2d_\mathrm{I}}{\lambda},\frac{2d_\mathrm{I}}{\lambda}\right]$ respectively denote the array phases along the $x$- and $y$-axis directions, and $d_\mathrm{I}$ denotes the spacing between any two adjacent IRS elements along the $x$/$y$-axis direction.

Based on the above, the IRS\textrightarrow BS channel $\boldsymbol{G}$ is modeled as a geometric multipath channel given by
\begin{equation}\label{irs-bs}
\boldsymbol{G} = 
\sum^L_{l=1} a_{l}  \boldsymbol{e}\left( \zeta_l,N_{\mathrm{B}} \right) \boldsymbol{u}^T\left( \vartheta_l,\psi_l \right),
\end{equation} 
where $L$ denotes the number of paths, $a_l \in \mathbb{C}$ denotes the complex-valued path gain of the $l$-th path, $\zeta_l = \frac{2d_\mathrm{B}}{\lambda} \cos {\theta}_{\mathrm{B},l}$ denotes the array phase of the BS with ${\theta}_{\mathrm{B},l}$ being the AoA of the $l$-th path,
and $\vartheta_l =\frac{2d_\mathrm{I}}{\lambda} \cos \theta_l \cos \phi_l$ and $\psi_l =\frac{2d_\mathrm{I}}{\lambda} \cos \theta_l \sin \phi_l $ denote the array phases of the IRS along the $x$- and $y$-axis directions, with $\theta_l$ and $\phi_l$ being the elevation and azimuth AoDs of the $l$-th path, respectively.


Due to the relatively short distance between the serving IRS and the user, the corresponding user\textrightarrow IRS channel $\boldsymbol{q}^{\left[n\right]}$ is modeled as a time-varying line-of-sight (LoS) channel given by
\begin{equation}\label{user-irs}
\boldsymbol{q}^{[n]} = {a}^{[n]}
{\boldsymbol{u}\left( {\vartheta}^{[n]},{\psi}^{[n]} \right)},
\end{equation} 
where ${a}^{[n]} \in \mathbb{C}$ denotes the complex-valued path gain in block $n$,
${\vartheta}^{[n]} = \frac{2d_\mathrm{I}}{\lambda} \cos {\theta}^{[n]} \cos {\phi}^{[n]}$ and ${\psi}^{[n]} = \frac{2d_\mathrm{I}}{\lambda} \cos {\theta}^{[n]} \sin {\phi}^{[n]}$ denote the array phases along the $x$- and $y$-axis directions, with ${\theta}^{[n]}$ and ${\phi}^{[n]}$ being the time-varying elevation and azimuth AoAs from the user to the IRS, respectively.

Let $\boldsymbol{\nu}^{[n]} = \left[\nu^{[n]}_{1},\ldots, \nu^{[n]}_{M}\right]^T$ denote the reflection coefficients of IRS in block $n$, where the reflection amplitudes of all reflecting elements are set to one or the maximum value to maximize the signal reflection power as well as ease the hardware implementation, thus leading to $\left|\nu^{[n]}_{m}\right| = 1$, $\forall m = 1,\ldots,M$.
Under the above setup, the IRS-reflected channel in block $n$ is given by
\begin{align}\label{i_reflected_cha}
\boldsymbol{h}^{[n]}_r 
= \boldsymbol{G} \operatorname{diag}\left(\boldsymbol{\nu}^{[n]}\right) \boldsymbol{q}^{[n]} = \underbrace{\boldsymbol{G} \operatorname{diag}\left(\boldsymbol{q}^{[n]}\right)}_{\boldsymbol{Q}^{[n]}} \boldsymbol{\nu}^{[n]},
\end{align} 
where $\boldsymbol{Q}^{[n]} \in \mathbb{C}^{N_{\mathrm{B}} \times M}$ denotes the cascaded user\textrightarrow IRS\textrightarrow BS channel in block $n$. Hence, the overall user\textrightarrow BS channel in block $n$ is given by
\begin{align}\label{eff_cha}
\boldsymbol{h}^{[n]}
= \boldsymbol{Q}^{[n]} \boldsymbol{\nu}^{[n]} +\boldsymbol{d}^{[n]}.
\end{align} 
\subsection{IRS Reflection Design under Perfect CSI}
\label{sec_refl}
For any given $\boldsymbol{\nu}^{[n]}$ in (\ref{eff_cha}), it is known that the maximum-ratio combining (MRC) is the optimal receive beamforming at the BS to maximize the received signal-to-noise ratio (SNR). Accordingly, the optimization problem for maximizing the overall user\textrightarrow BS channel power gain in (\ref{eff_cha}) is formulated as follows.
\begin{align}\label{eff_cg}
\text { (P1): } \max_{\boldsymbol{\nu}^{[n]}} \quad & \left\|\boldsymbol{Q}^{[n]} \boldsymbol{\nu}^{[n]}  + \boldsymbol{d}^{[n]}\right\|^2 \nonumber\\
\textrm{s.t.} \quad & \left|\nu^{[n]}_{m} \right|= 1, \quad \forall m=1, \ldots, M. 
\end{align}
It can be verified that (P1) is a non-convex optimization problem, which only admits a closed-form optimal solution when $\operatorname{rank}\left(\boldsymbol{G}\right) = 1$ and/or $N_{\mathrm{B}} = 1$ \cite{irs_tut}. However, in other general setups, it is difficult to perfectly align $\boldsymbol{Q}^{[n]}$ and $\boldsymbol{d}^{[n]}$ in the objective function of (P1) by tuning $\boldsymbol{\nu}^{[n]}$, due to the unit-modulus constraints given in (\ref{eff_cg}), even if $M\ge N_\mathrm{B}$.
Given the perfect CSI of $\boldsymbol{Q}^{[n]}$ and $\boldsymbol{d}^{[n]}$, various methods have been developed in the literature to obtain high-quality suboptimal solutions for (P1), such as 1) semidefinite relaxation (SDR) with Guassian randomization \cite{irs1}, and 2) alternating optimization (AO) where each of the phase shifts $\boldsymbol{\nu}^{[n]}_{m}, m=1, \ldots, M$, is alternately optimized in closed-form as in \cite{ao_pass} with the others being fixed in an iterative manner, which only guarantees the convergence to a locally optimal solution in general.
\subsection{Main Issues for Implementing IRS-Aided High-Mobility Communication} \label{sec_chal}
However, the acquisition of the full CSI of $\boldsymbol{Q}^{[n]}$ in addition to the direct channel $\boldsymbol{d}^{[n]}$ in each transmission block $n$ is practically challenging for the IRS-aided high-mobility communication, as elaborated in the following.
\begin{enumerate}
\item \textbf{Modification of Existing Transmission Protocol:} 
Dedicated pilot symbols are required for the IRS to adjust its reflection over time to facilitate the channel estimation of $\boldsymbol{Q}^{[n]}$ at the BS, in addition to estimating the direct channel $\boldsymbol{d}^{[n]}$ (see, e.g., \cite{liu, guan,sb1,dll,sb2}), which needs to modify the existing uplink transmission protocol (e.g., that of massive MIMO communication) which only estimates the direct channel $\boldsymbol{d}^{[n]}$ without the involvement of IRS.

\item \textbf{Prohibitive Channel Estimation Overhead Due to High Mobility:} 
Even if both $\boldsymbol{Q}^{[n]}$ and $\boldsymbol{d}^{[n]}$ can be estimated at the BS by modifying the existing transmission protocol \cite{liu, guan,dll}, the required additional training overhead will be prohibitive as it is generally proportional to the number of reflecting elements of the IRS, $M$, or the number of IRS sub-surfaces each constituting a number of adjacent reflecting elements by applying the element grouping strategy in \cite{ofdm1,grouping1}.
Considering the short block duration which is typical in the high-mobility communication scenario, the time left for data transmission in each block will be severely reduced, causing a significant rate loss that may even overwhelm the IRS beamforming/SNR gain and thus result in even lower communication throughput as compared to that of the conventional system without IRS.

\item \textbf{Non-Negligible Real-Time Feedback Delay:} 
Based on the CSI of $\boldsymbol{Q}^{[n]}$ and $\boldsymbol{d}^{[n]}$ estimated at the BS, the passive beamforming vector $\boldsymbol{\nu}^{[n]}$ is designed and then fed back to the IRSC to adjust its IRS reflection \cite{ofdm1,grouping1}, which inevitably incurs feedback delay that may render the designed IRS reflection outdated and less effective as the user-IRS channel varies rapidly due to the user's high mobility.


\end{enumerate}

\section{Proposed Passive Beamforming Design} \label{sec_bfpro}
As mentioned in Section \ref{sec_refl}, it is difficult to perfectly align the IRS-reflected channel with the direct channel under the general setup with the multi-antenna BS. In fact, with a large $M$, the performance gain achieved by aligning the IRS-reflected channel with the direct channel becomes marginal as the power of the IRS-reflected channel becomes dominant over that of the direct channel.
On the other hand, it is also practically challenging to acquire the real-time CSI of both $\boldsymbol{Q}^{[n]}$ and $\boldsymbol{d}^{[n]}$ to perfectly align them in the high-mobility communication scenario. 
Motivated by the above, instead of maximizing the overall user\textrightarrow BS channel gain as in (P1), we propose to design the IRS reflection to maximize the IRS-reflected channel gain only based on the CSI of $\boldsymbol{Q}^{[n]}$, as shown next.

With the direct channel omitted, (P1) is reduced to the following optimization problem for maximizing the power of the IRS-reflected channel given in (\ref{i_reflected_cha}).
\begin{align}
\text { (P2): } \max_{\boldsymbol{\nu}^{[n]}} \quad & \left\|\boldsymbol{Q}^{[n]} \boldsymbol{\nu}^{[n]}\right\|^2 \nonumber\\
\textrm{s.t.} \quad & \left|\nu^{[n]}_{m} \right|= 1, \quad \forall m=1, \ldots, M. \label{p_cons}
\end{align}
Although (P2) is still a non-convex optimization problem, it can be sub-optimally solved via the SDR \cite{irs1} for each block $n$ based on the cascaded CSI $\boldsymbol{Q}^{[n]}$.
However, the SDR algorithm generally requires a large number of iterations to reach convergence and also has a relatively high computational complexity in the order of $\mathcal{O}\left(M^{4.5}\right)$ for each iteration, which may not be implementable in real time for the high-mobility vehicular communication. 
Fortunately, we notice from (\ref{i_reflected_cha}) that the cascaded channel $\boldsymbol{Q}^{[n]}$ can be decomposed into the inner product of the time-invariant IRS\textrightarrow BS channel $\boldsymbol{G}$ and the time-varying user\textrightarrow IRS channel $\boldsymbol{q}^{[n]}$. Based on this decomposed structure, we propose to first obtain an initial passive beamforming vector based on the knowledge of $\boldsymbol{G}$. 
Then, with the CSI of $\boldsymbol{q}^{[n]}$ given in (\ref{user-irs}) available, we can further design the real-time passive beamforming vector as
\begin{align}\label{online_pbf}
\boldsymbol{\nu}^{[n]} = \operatorname{diag}\left(\boldsymbol{u}^*\left(\vartheta^{[n]},\psi^{[n]}\right)\right)\boldsymbol{\bar{\nu}},
\end{align}
where $\boldsymbol{\bar{\nu}} = \left[\bar{\nu}_1,\ldots, \bar{\nu}_M\right]^T$ is the initial passive beamforming vector with $\left|\bar{\nu}_{m} \right|= 1, \forall m=1, \ldots, M$. Accordingly, by substituting (\ref{online_pbf}) into (P2), we have
\begin{align}
\text { (P3): } \max_{\boldsymbol{\bar{\nu}}} \quad & \left\|\boldsymbol{G} \boldsymbol{\bar{\nu}}\right\|^2 \nonumber\\
\textrm{s.t.} \quad & \left|\bar{\nu}_{m} \right|= 1, \quad \forall m=1, \ldots, M, \label{p22_cons}
\end{align}
where the objective function in (P3) is obtained since we have
\begin{align}
\left\|\boldsymbol{Q}^{[n]} \boldsymbol{\nu}^{[n]}\right\|^2 &= \left\|\boldsymbol{G} \operatorname{diag}\left(\boldsymbol{q}^{[n]}\right) \operatorname{diag}\left(\boldsymbol{u}^*\left(\vartheta^{[n]},\psi^{[n]}\right)\right) \boldsymbol{\bar{\nu}}\right\|^2 = \left\| {a}^{[n]} \boldsymbol{G} \boldsymbol{\bar{\nu}}\right\|^2.
\end{align}
As such, given the static CSI of $\boldsymbol{G}$, the initial passive beamforming vector can be designed off-line via the SDR algorithm by solving (P3) as $\boldsymbol{\bar{\nu}}_{\mathrm{ini}}$. Next, with the real-time CSI in terms of the angle/phase information on $\vartheta^{[n]}$ and $\psi^{[n]}$ of the user\textrightarrow IRS channel $\boldsymbol{q}^{[n]}$ in (\ref{user-irs}) acquired at the IRSC in each block $n$ (to be shown in Section \ref{sec_ce}), the real-time passive beamforming vector for the IRS can be set according to (\ref{online_pbf}) by its IRSC (i.e., without the need of BS's real-time feedback).

\section{New Channel Estimation Scheme} \label{sec_ce}
\begin{figure*}
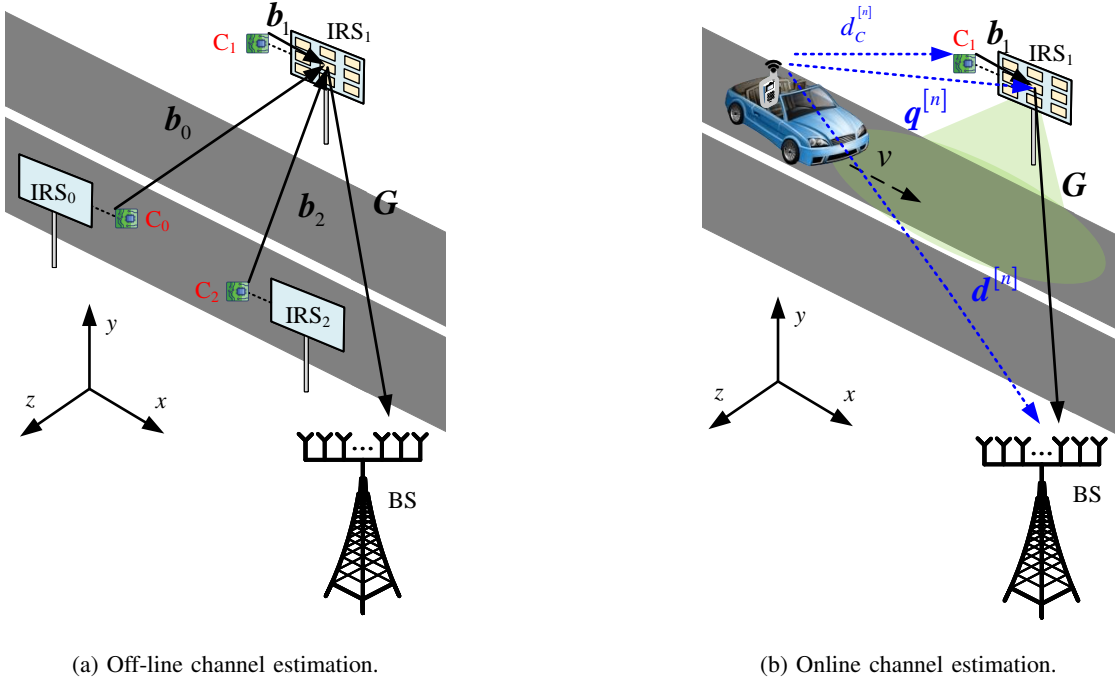
 
\centering
     \begin{subfigure}[t]{0.45\textwidth}
         \centering
         \includegraphics{config_off.pdf}
         \caption{Off-line channel estimation.}
     \end{subfigure}
     \hfill
     \begin{subfigure}[t]{0.45\textwidth}
         \centering
         \includegraphics{config_on.pdf}
         \caption{Online channel estimation.}
     \end{subfigure}
     \caption{Illustration of the proposed channel estimation scheme.}
     \label{config_esti}
\end{figure*}

For the IRS reflection design proposed in Section \ref{sec_bfpro}, we only need to acquire the static IRS 1\textrightarrow BS channel $\boldsymbol{G}$ off-line and the angle/phase information $\{\vartheta^{[n]},\psi^{[n]}\}$ of the time-varying user\textrightarrow IRS channel $\boldsymbol{q}^{[n]}$ in real time. 
To this end, we propose a new scheme in this section to realize the above channel estimation procedures at the BS and IRSC 1, respectively, in a decentralized manner. The proposed channel estimation consists of off-line and online stages as shown in  Fig.~\ref{config_esti}, outlined as follows.
\begin{enumerate}
\item \textbf{Off-line Channel Estimation and Initial Passive Beamforming:} 
As shown in Fig.~\ref{config_esti}(a), IRSC $k$ (labeled as C\textsubscript{\textit{k}}) with $k \in \{0,2\}$, i.e., the controllers of the two IRSs nearest the serving IRS (i.e., IRS 1, labeled as IRS\textsubscript{1}) and symmetrically located at its opposite roadside, transmit pilot signals to the BS consecutively for estimating a scaled IRS\textsubscript{1}-BS channel $\boldsymbol{\bar{G}}$, i.e., $\boldsymbol{\bar{G}} = \dot{a}_0 \boldsymbol{G}$ with $\dot{a}_0$ denoting a scaling factor. 
Based on the estimated $\boldsymbol{\bar{G}}$, BS computes the initial passive beamforming vector $\boldsymbol{\bar{\nu}}_{\mathrm{ini}}$ by solving (P3) and feeds it back to IRSC 1 (labeled as C\textsubscript{1}).
Next, C\textsubscript{1} transmits pilot signals to the BS for estimating a scaled IRS\textsubscript{1}\textrightarrow C\textsubscript{1} channel $\boldsymbol{\bar{b}}_1$, where $\boldsymbol{\bar{b}}_1 = \frac{\boldsymbol{b}_1}{\dot{a}_0} $ and is fed back to C\textsubscript{1} as well. 
\item \textbf{Online Channel Estimation and Passive Beam Prediction:} 
As shown in Fig.~\ref{config_esti}(b), before IRS\textsubscript{1} starts to serve the moving user (i.e., when $n \leq 0$), C\textsubscript{1} estimates the real-time angle/phase information $\{\vartheta^{[n]},\psi^{[n]}\}$ at each block $n$, by exploiting the pilot signals transmitted by the user in each block to the BS as in the existing uplink transmission protocol.
When the estimated $\{\hat{\vartheta}^{[n]},\hat{\psi}^{[n]}\}$ reaches a pre-defined coverage region of $\mathrm{IRS}_1$, C\textsubscript{1} sets $n = 0$ and starts predicting $\{\hat{\vartheta}^{[n]},\hat{\psi}^{[n]}\}$ for each block of $n >0$ based on the previously estimated information $\{\hat{\vartheta}^{[n]},\hat{\psi}^{[n]}\}_{n\leq 0}$. With $\boldsymbol{\bar{\nu}}_{\mathrm{ini}}$ computed off-line and the predicted information on $\{\hat{\vartheta}^{[n]},\hat{\psi}^{[n]}\}$, C\textsubscript{1} dynamically adjusts the passive beamforming of IRS\textsubscript{1} for each block $n$ according to (\ref{online_pbf}).  
\end{enumerate}
\begin{remark}
With $\boldsymbol{\bar{G}}$ and $\boldsymbol{\bar{b}}_1$ estimated off-line, C\textsubscript{1} only needs to estimate/predict $\{\vartheta^{[n]},\psi^{[n]}\}_{n>0}$ based on the pilot signals transmitted by the user in each block $n$ to the BS, which can be achieved efficiently without modifying the existing uplink transmission protocol. 
With the predicted information on $\{\hat{\vartheta}^{[n]},\hat{\psi}^{[n]}\}$ for $n>0$, the passive beamforming vector for IRS\textsubscript{1} can be independently set by C\textsubscript{1} in real time (i.e., without the need of the BS's real-time feedback), which also avoids the handover between adjacent IRSs as it incurs additional signaling overhead. As a result, the main issues for implementing the IRS-aided high-mobility communication mentioned in Section \ref{sec_chal} are all tackled.
\end{remark}
\begin{figure} 
\centering
     \begin{subfigure}[t]{0.7\textwidth}
         \centering
         \includegraphics[width=\textwidth]{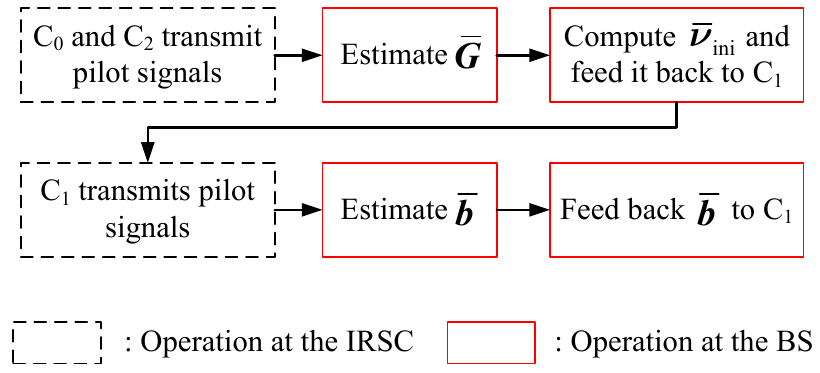}
         \caption{Flow chart of the off-line stage.}
     \end{subfigure}
     \begin{subfigure}[t]{0.7\textwidth}
         \centering
         \includegraphics[width=\textwidth]{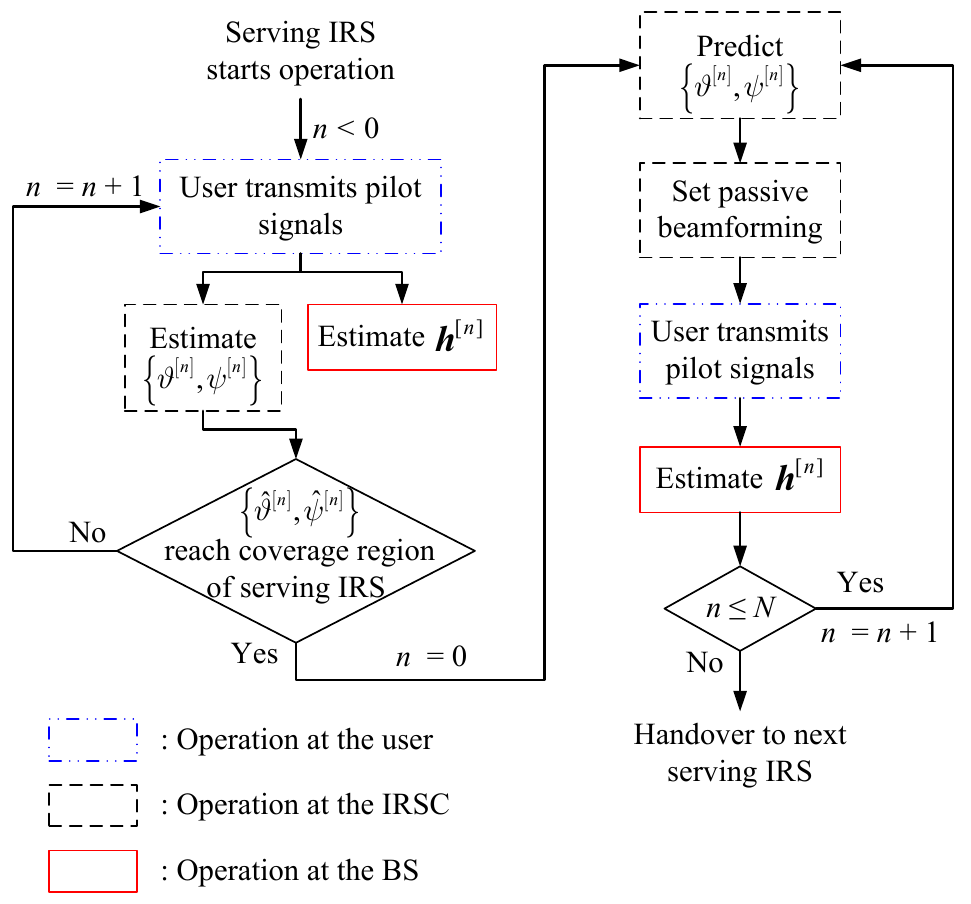}
         \caption{Flow chart of the online stage.}
     \end{subfigure}
     \caption{Illustration of the proposed IRS operating protocols.}
     \label{fc_all}
\end{figure}

To summarize the proposed IRS operating protocols for assisting the high-speed vehicular communication, we illustrate them in Figs.~\ref{fc_all}(a) and \ref{fc_all}(b) for the off-line and online stages, respectively.

\subsection{Off-line Channel Estimation and Initial Passive Beamforming}
In this subsection, we present the details of the proposed off-line channel estimation.
The direct link between the BS and C\textsubscript{\textit{k}}, $k \in \{0,1,2\}$, is omitted for brevity\footnote{The direct link can be estimated at the BS using the conventional channel estimation scheme with the IRS turned OFF or via the estimation scheme in \cite{ofdm2} with the IRS turned ON.}. 
As shown in Fig.~\ref{config_esti}(a), we denote the static C\textsubscript{\textit{k}}\textrightarrow IRS\textsubscript{1} channel by $\boldsymbol{b}_k\in \mathbb{C}^{M\times 1}$, with $k \in \{0,1,2\}$.
Hence, the cascaded C\textsubscript{\textit{k}}\textrightarrow IRS\textsubscript{1}\textrightarrow BS channel is represented by $\boldsymbol{R}_k = \boldsymbol{G}\operatorname{diag}\left(\boldsymbol{b}_k\right)$, $k \in \{0 ,1,2\}$.
Due to the relatively short distance between C\textsubscript{\textit{k}} and IRS\textsubscript{1} for $k \in \{0,2\}$, the corresponding C\textsubscript{\textit{k}}\textrightarrow IRS\textsubscript{1} channel $\boldsymbol{b}_k$ is modeled as the far-field LoS channel given by
\begin{equation}\label{airsc_irs}
\boldsymbol{b}_{k}= \dot{a}_k {\boldsymbol{u}\left( \dot{\vartheta}_k,\dot{\psi}_k \right)}, \quad k \in \{0,2\},
\end{equation}
where $\dot{a}_k \in \mathbb{C}$ denotes the path gain, $\dot{\vartheta}_k = \frac{2d_\mathrm{I}}{\lambda} \cos \dot{\theta}_k \cos \dot{\phi}_k$ and
$\dot{\psi}_k = \frac{2d_\mathrm{I}}{\lambda} \cos \dot{\theta}_k \sin \dot{\phi}_k$ denote the array phases of IRS\textsubscript{1} along the $x$- and $y$-axis directions, with $\dot{\theta}_k$ and $\dot{\phi}_k$ being the elevation and azimuth AoAs from C\textsubscript{\textit{k}} to IRS\textsubscript{1}, respectively. As shown in Fig.~\ref{config_esti}(a), we assume that the placements of C\textsubscript{0} and C\textsubscript{2} are {\it symmetrical} (with respect to IRS\textsubscript{1}) with the same elevation AoAs (i.e., $\dot{\phi}_0 = 0$, $\dot{\phi}_2 = \pi$, and $\dot{\theta}_0 = \dot{\theta}_2$), such that we have $\dot{\vartheta}_0 = - \dot{\vartheta}_2$ and $\dot{\psi}_0 = \dot{\psi}_2 = 0$, which will be utilized for facilitating the off-line channel estimation in the following.

\subsubsection{Step 1: Estimation of $\boldsymbol{\bar{G}}$.} \label{sec_o1}
Let $M_0$ denote the number of pilot symbols transmitted by C\textsubscript{0} and $\mathcal{M}_0 \triangleq \{1,\ldots,M_0\}$ denote the index set.
With the pilot symbol $x^{(i)}_{0} = 1$ transmitted by C\textsubscript{0}, the received signal vector at the BS is written as
\begin{align}\label{off_1_sig}
\boldsymbol{y}^{\left(i\right)}_{{0}}
 &= \underbrace{\boldsymbol{G} \operatorname{diag}\left(\boldsymbol{b}_{0}\right)}_{\boldsymbol{R}_{0}}
\boldsymbol{{\nu}}^{\left(i\right)}_{{0}} + \boldsymbol{z}^{\left(i\right)}_{{0}}, \quad i \in \mathcal{M}_0,
\end{align} 
where $\boldsymbol{{\nu}}^{\left(i\right)}_{{0}} \in \mathbb{C}^{M \times 1}$ denotes the training reflection vector of IRS\textsubscript{1}, $\boldsymbol{z}^{\left(i\right)}_{{0}} \sim\mathcal{N}_{c}\left(\boldsymbol{0}, \sigma^{2} \boldsymbol{\mathrm{I}}_{N_{\mathrm{B}}}\right)$ denotes the additive white Gaussian noise (AWGN) vector at the BS with $\sigma^{2}$ being the normalized noise power. By stacking $M_0$ received pilot vectors $\{\boldsymbol{y}^{\left(i\right)}_{{0}}\} _{i\in\mathcal{M}_0}$ into $\boldsymbol{Y}_{{0}}=\left[\boldsymbol{y}^{\left(1\right)}_{{0}},\ldots,\boldsymbol{y}^{\left(M_0\right)}_{{0}}\right] \in \mathbb{C}^{N_\mathrm{B} \times M_0}$, we have
\begin{align}\label{off_1_sig_col}
\boldsymbol{Y}_{{0}}
 &=  \boldsymbol{R}_{0} \boldsymbol{{V}}_{{0}} + \boldsymbol{Z}_{{0}} = \boldsymbol{G} \operatorname{diag}\left(\boldsymbol{b}_{0}\right) \boldsymbol{{V}}_{{0}} + \boldsymbol{Z}_{{0}} \nonumber\\
&= \dot{a}_0\left(\sum^L_{l=1} a_l  \boldsymbol{e}\left( \zeta_l,N_{\mathrm{B}} \right) \boldsymbol{u}^T\left( \vartheta_l,\psi_l \right)\right) \operatorname{diag}\left({\boldsymbol{u}\left( \dot{\vartheta}_0,\dot{\psi}_0 \right)}\right)\boldsymbol{{V}}_{{0}}+ \boldsymbol{Z}_{{0}} \nonumber\\
&=\left(\sum^L_{l=1} a_{l,0}  \boldsymbol{e}\left( \zeta_l,N_{\mathrm{B}} \right) \boldsymbol{u}^T\left( {\vartheta}_{l,0},{\psi}_{l,0} \right)\right)
\boldsymbol{{V}}_{{0}}+ \boldsymbol{Z}_{{0}},
\end{align} 
where $a_{l,0} =  \dot{a}_0 a_l$ denotes the product path gain of the $l$-th path, $\vartheta_{l,0} = \vartheta_l+ \dot{\vartheta}_0$ and $\psi_{l,0} = \psi_l+ \dot{\psi}_0$ denote the effective array phases of IRS\textsubscript{1} along the $x$- and $y$-axis directions of the $l$-th path, respectively,
$\boldsymbol{{V}}_{{0}} =  \left[\boldsymbol{{\nu}}_{{0}}^{\left(1\right)}, \ldots, \boldsymbol{{\nu}}_{{0}}^{\left(M_0\right)} \right]\in \mathbb{C}^{M \times M_0}$ denotes the training reflection matrix at IRS\textsubscript{1}, $\boldsymbol{Z}_{{0}} = \left[\boldsymbol{z}_{{0}}^{\left(1\right)}, \ldots, \boldsymbol{z}_{{0}}^{\left(M_0\right)} \right]$ denotes the corresponding AWGN matrix at the BS.
For notational convenience, we define $\boldsymbol{\xi} = \left[a_1,\zeta_1,\vartheta_{1,0},{\psi}_{1,0},\ldots, a_L,\zeta_L,\vartheta_{L,0},{\psi}_{L,0}\right]^T$ as the collection of the unknown parameters in (\ref{off_1_sig_col}).
Based on (\ref{off_1_sig_col}), the maximum likelihood (ML) estimation of $\boldsymbol{\xi}$ at the BS is given by
\begin{align}\label{mle_off1}
&\{\boldsymbol{\hat{\xi}}\} = \arg \min _{\boldsymbol{\xi}} \left\|\boldsymbol{Y}_{{0}} - \left(\sum^L_{l=1} a_{l,0}  \boldsymbol{e}\left( \zeta_l,N_{\mathrm{B}} \right) \boldsymbol{u}^T\left( {\vartheta}_{l,0},{\psi}_{l,0} \right)\right)
\boldsymbol{{V}}_{{0}}\right\|^2_2.
\end{align}
However, the ML estimation given in (\ref{mle_off1}) incurs prohibitively high computational complexity due to the joint search over the high dimensional vector $\boldsymbol{\xi}$. 
To tackle this difficulty, we propose a low-complexity decoupled estimation scheme by first estimating the array phases and then the path gains successively. Specifically, we right multiply the pseudo inverse of $\boldsymbol{{V}}_{{0}}$ and perform the vectorization, yielding 
\begin{align}\label{off_1_sig_col_re}
\boldsymbol{\Tilde{y}}_{{0}}
&= \operatorname{vec}\left(\boldsymbol{Y}_{{0}} \boldsymbol{{V}}_{{0}}^\dagger \right) \nonumber\\
&= \operatorname{vec}\left(\sum^L_{l=1} a_{l,0}  \boldsymbol{e}\left( \zeta_l,N_{\mathrm{B}} \right) \boldsymbol{u}^T\left( {\vartheta}_{l,0},{\psi}_{l,0} \right)\right) +  \operatorname{vec}\left(\boldsymbol{Z}_{{0}} \boldsymbol{{V}}_{{0}}^\dagger \right) \nonumber\\
&=
\underbrace{\left[  \boldsymbol{u}\left( {\vartheta}_{1,0},{\psi}_{1,0} \right)\otimes \boldsymbol{e}\left( \zeta_1,N_{\mathrm{B}} \right), \ldots,  \boldsymbol{u}\left( {\vartheta}_{l,0},{\psi}_{l,0} \right) \otimes \boldsymbol{e}\left( \zeta_L,N_{\mathrm{B}} \right) \right]}_{\boldsymbol{W} \left( \boldsymbol{\mu}_0 \right)} 
\boldsymbol{a}_0
+  \operatorname{vec}\left(\boldsymbol{Z}_{{0}} \boldsymbol{{V}}_{{0}}^\dagger \right) ,
\end{align} 
where $\boldsymbol{\mu}_0= \left[\zeta_1,\vartheta_{1,0},{\psi}_{1,0},\ldots,\zeta_L,\vartheta_{L,0},{\psi}_{L,0}\right]^T$ denotes the collection of the array phases, $\boldsymbol{W} \left( \boldsymbol{\mu}_0 \right)\in \mathbb{C}^{N_{\mathrm{B}} M\times L}$ denotes the effective array response matrix,
and $\boldsymbol{a}_0 = \left[a_{1,0}, \ldots, a_{L,0}\right]^T$ denotes the path gain vector. 
Note that we need to properly construct $\boldsymbol{{V}}_{{0}}$ such that $M_0\geq \operatorname{rank}\left(\boldsymbol{{V}}_{{0}}\right) \geq M$ to ensure the existence of the pseudo inverse of $\boldsymbol{{V}}_{{0}}$. 
In practice, we can apply the $M\times M$ discrete Fourier transform (DFT) matrix for designing $\boldsymbol{{V}}_{{0}}$. 
According to (\ref{off_1_sig_col_re}), existing radio direction finding algorithms such as multiple signal classification (MUSIC) \cite{music} can be applied to estimate the array phases $\boldsymbol{\mu}_0$. With the estimated $\boldsymbol{\hat{\mu}}_0$, the least squares (LS) estimate of the path gain vector $\boldsymbol{a}_0$ is given by
\begin{align}\label{ls_pg}
\boldsymbol{\hat{a}}_0 = \left(\boldsymbol{W} \left( \boldsymbol{\hat{\mu}}_0 \right)\right)^\dagger \boldsymbol{\Tilde{y}}_{{0}}.
\end{align} 

Following the similar procedures as above, the array phases with respect to the C\textsubscript{2}\textrightarrow IRS\textsubscript{1}\textrightarrow BS channel (i.e, $\{\vartheta_{1,2},{\psi}_{1,2},\ldots, \vartheta_{L,2},{\psi}_{L,2}\}$) can be estimated. By exploiting the symmetrical placement of C\textsubscript{0} and C\textsubscript{2} (i.e., $\dot{\vartheta}_0 = - \dot{\vartheta}_2$ and $\dot{\psi}_0 = \dot{\psi}_2=0$), the estimates of the array phases of the IRS\textsubscript{1}\textrightarrow BS channel $\boldsymbol{G}$ are given by
\begin{align}\label{off_effx}
\hat{{\vartheta}}_{l} = \frac{\hat{{\vartheta}}_{l,0}+\hat{{\vartheta}}_{l,2}}{2}, \quad l \in \{1,\ldots, L\},
\end{align} 
\begin{align}\label{off_effy}
\hat{{\psi}}_{l} = \frac{\hat{{\psi}}_{l,0}+\hat{{\psi}}_{l,2}}{2} , \quad l \in \{1,\ldots, L\},
\end{align} 
regardless of the path gains in $\boldsymbol{G}$ (i.e., $\left[a_1,\ldots,a_L\right]$). Assuming perfect estimation  in (\ref{ls_pg})-(\ref{off_effy}), we can reconstruct the scaled IRS\textsubscript{1}\textrightarrow BS channel as 
\begin{align}\label{app_g}
\bar{\boldsymbol{G}}
&= \sum^L_{l=1} \underbrace{\dot{a}_0 a_l }_{a_{l,0}}  \boldsymbol{e}\left( \zeta_l,N_{\mathrm{B}} \right) \boldsymbol{u}^T\left( \vartheta_l,\psi_l \right),
\end{align} 
where $\bar{\boldsymbol{G}} = \dot{a}_0 \boldsymbol{G}$ can be regarded as a scaled version of $\boldsymbol{G}$. 
It can be verified that by replacing $\boldsymbol{G}$ with $\bar{\boldsymbol{G}}$ in (P3), the scaling factor $\dot{a}_0$ in $\bar{\boldsymbol{G}}$ will not affect the design of $\boldsymbol{\bar{\nu}}_{\mathrm{ini}}$.
In practice, the estimate of $\bar{\boldsymbol{G}}$ can be calculated by substituting (\ref{ls_pg})-(\ref{off_effy}) into (\ref{app_g}).
After that, the BS designs $\boldsymbol{\bar{\nu}}_{\mathrm{ini}}$ by solving (P3) with $\boldsymbol{G}$ replaced by the estimate of $\bar{\boldsymbol{G}}$ and then feeds it back to C\textsubscript{1}.

\subsubsection{Step 2: Estimation of $\boldsymbol{\bar{b}}_1$.}
Let $M_1$ denote the number of pilot symbols transmitted by C\textsubscript{1} and $\mathcal{M}_1 \triangleq \{1,\ldots,M_1\}$ denote the index set.
With the pilot symbol $x^{(i)}_{1} = 1$ transmitted by C\textsubscript{1}, the received signal vector at BS is written as
\begin{align}\label{off_2_sig}
\boldsymbol{y}^{\left(i\right)}_{{1}}
 &= \underbrace{\boldsymbol{G} \operatorname{diag}\left(\boldsymbol{b}_1\right)}_{\boldsymbol{R}_1}
\boldsymbol{{\nu}}^{\left(i\right)}_{{1}}  + \boldsymbol{z}^{\left(i\right)}_{{1}}\nonumber\\
&= \underbrace{\frac{1}{\dot{a}_0} \bar{\boldsymbol{G}} \operatorname{diag}\left(\boldsymbol{b}_1\right)}_{\boldsymbol{R}_1}
\boldsymbol{{\nu}}^{\left(i\right)}_{{1}}  + \boldsymbol{z}^{\left(i\right)}_{{1}}
, \quad i \in \mathcal{M}_1,
\end{align} 
where $\boldsymbol{{\nu}}^{\left(i\right)}_{{1}} \in \mathbb{C}^{M \times 1}$ denotes the training reflection vector of IRS\textsubscript{1}, $\boldsymbol{z}^{\left(i\right)}_{{1}} \sim\mathcal{N}_{c}\left(\boldsymbol{0}, \sigma^{2} \boldsymbol{\mathrm{I}}_{N_{\mathrm{B}}}\right)$ denotes the AWGN vector at the BS. 
By stacking $M_1$ received pilot vectors $\{\boldsymbol{y}^{\left(i\right)}_{{1}}\} _{i\in\mathcal{M}_1}$ into $\boldsymbol{Y}_{{1}}=\left[\boldsymbol{y}^{\left(1\right)}_{{1}},\ldots,\boldsymbol{y}^{\left(M_1\right)}_{{1}}\right] \in \mathbb{C}^{N_\mathrm{B} \times M_1}$, we have
\begin{align}\label{off_2_sig_coll}
\boldsymbol{Y}_{{1}}
 &= \underbrace{\frac{1}{\dot{a}_0} \bar{\boldsymbol{G}} \operatorname{diag}\left(\boldsymbol{b}_1\right)}_{\boldsymbol{R}_1}
\boldsymbol{{V}}_{{1}} + \boldsymbol{Z}_{{1}}, 
\end{align} 
where $\boldsymbol{{V}}_{{1}} =  \left[\boldsymbol{{\nu}}_{{1}}^{\left(1\right)}, \ldots, \boldsymbol{{\nu}}_{{1}}^{\left(M_1\right)} \right]\in \mathbb{C}^{M \times M_1}$ denotes the training reflection matrix at IRS\textsubscript{1}, $\boldsymbol{Z}_{{1}} =  \left[\boldsymbol{z}_{{1}}^{\left(1\right)}, \ldots, \boldsymbol{z}_{{1}}^{\left(M_1\right)} \right]$ denotes the corresponding AWGN matrix at the BS. By properly constructing $\boldsymbol{{V}}_{{1}}$ such that $M_1\geq\operatorname{rank}\left(\boldsymbol{{V}}_{{1}}\right) \geq M$, $\boldsymbol{R}_1$ can be estimated as 
\begin{equation}\label{BB}
\hat{\boldsymbol{R}}_1 = \boldsymbol{Y}_{{1}} \boldsymbol{{V}}_{{1}}^\dagger = \underbrace{\frac{1}{\dot{a}_0} \bar{\boldsymbol{G}} \operatorname{diag}\left(\boldsymbol{b}_1\right)}_{\boldsymbol{R}_1}  + \boldsymbol{Z}_{{1}} \boldsymbol{{V}}_{{1}}^\dagger.
\end{equation}
Based on $\boldsymbol{R}_1$ and $\bar{\boldsymbol{G}}$, we can reconstruct a scaled version of $\boldsymbol{b}_1$ as
\begin{equation}\label{est_b}
\bar{\boldsymbol{b}}_1 = \frac{1}{N_{\mathrm{B}}} \sum_{n=1}^{N_{\mathrm{B}}} \left[\boldsymbol{{R}}_1^T \oslash \boldsymbol{\bar{G}}^T\right]_{:,n},
\end{equation}
where $\oslash$ denotes the element-wise division and we have $\boldsymbol{\bar{b}}_1 = \frac{1}{\dot{a}_0} \boldsymbol{b}_1$. It can be verified that the scaling factor $\frac{1}{\dot{a}_0} $ in $\boldsymbol{\bar{b}}_1$ will not affect the estimation of the angle/phase information $\{\vartheta^{[n]}, \psi^{[n]}\}$ and thus the passive beamforming design in (\ref{online_pbf}).
In practice, the estimate of $\bar{\boldsymbol{b}}_1$ can be calculated by substituting the estimate of $\bar{\boldsymbol{G}}$ and (\ref{BB}) into (\ref{est_b}).
After that, the BS feeds back the estimate of $\bar{\boldsymbol{b}}_1$ to C\textsubscript{1}.

\subsection{Online Channel Estimation and Passive Beam Prediction}
In the conventional uplink transmission protocol, the user transmits $\tau$ pilot symbols for the BS to estimate the user\textrightarrow BS channel during each block.
As shown in Fig.~\ref{config_esti}(b), we denote the time-varying channel of the direct user\textrightarrow C\textsubscript{1} link by $d_\mathrm{C}^{[n]} \in \mathbb{C}$ in block $n$.
Let $\mathcal{T} \triangleq \{1, \ldots, \tau \}$ denote the index set for the $\tau$ pilot symbols.

\subsubsection{Estimation of $\vartheta^{[n]}$ and $\psi^{[n]}$ for $n \leq 0$ (when IRS\textsubscript{1} has not served the user yet).} \label{sec_onest}
With $x^{(i)}_\mathrm{P} = 1$ being the pilot symbol, the received signal at C\textsubscript{1} during time slot $i$ of block $n$ can be written as 
\begin{align}\label{on_ef_sig}
y^{[n],\left(i\right)}
 &= \left(\boldsymbol{\nu}^{\left(i\right)}\right)^T \operatorname{diag}\left(\boldsymbol{b}_1 \right) \boldsymbol{q}^{[n]} + d^{[n]}_{\mathrm{C}} + z^{[n],\left(i\right)} \nonumber\\
 &= a^{[n]} \left(\boldsymbol{\nu}^{\left(i\right)}\right)^T \operatorname{diag}\left(\boldsymbol{b}_1 \right) \boldsymbol{u}\left({\vartheta^{[n]}},\psi^{[n]}\right) + d^{[n]}_{\mathrm{C}} + z^{[n],\left(i\right)} \nonumber\\
 &= \underbrace{\dot{a}_0 a^{[n]}}_{\bar{a}^{[n]}} \left(\boldsymbol{\nu}^{\left(i\right)}\right)^T \operatorname{diag}\left(\boldsymbol{\bar{b}}_1 \right) \boldsymbol{u}\left({\vartheta^{[n]}},\psi^{[n]}\right) + d^{[n]}_{\mathrm{C}} + z^{[n],\left(i\right)}
 , \quad \forall i \in \mathcal{T},
\end{align}
where $\boldsymbol{\nu}^{\left(i\right)} \in \mathbb{C}^{M \times 1}$ denotes the training reflection vector of IRS\textsubscript{1}, 
$z^{[n],\left(i\right)} \sim\mathcal{N}_{c}\left(0, \sigma^{2}\right)$ denotes the receiver noise at C\textsubscript{1}, and $\bar{a}^{[n]} = \dot{a}_0 a^{[n]}$ denotes the effective path gain.
By stacking $\tau$ received pilot symbols $\{y^{[n],\left(i\right)}\}_{i\in \mathcal{T}}$ into $\boldsymbol{y}^{[n]} = \left[y^{[n],\left(1\right)},\ldots,y^{[n],\left(\tau\right)}  \right]^T \in \mathbb{C}^{\tau \times 1}$, we have
\begin{equation}\label{on_sigs}
\boldsymbol{y}^{[n]}
 =  \bar{a}^{[n]} \boldsymbol{V} \operatorname{diag}\left(\boldsymbol{\bar{b}}_1 \right)
\boldsymbol{u}\left({\vartheta^{[n]}},\psi^{[n]}\right) +  d^{[n]}_{\mathrm{C}} \boldsymbol{1}_{\tau }+ \boldsymbol{z}^{[n]}, 
\end{equation} 
where $\boldsymbol{V} = \left[\boldsymbol{\nu}^{\left(1\right)}, \ldots, \boldsymbol{\nu}^{\left(\tau\right)} \right]^T \in \mathbb{C}^{\tau \times M}$ denotes the training reflection matrix at IRS\textsubscript{1} and $\boldsymbol{z}^{[n]} = \left[z^{[n],\left(1\right)},\ldots,z^{[n],\left(\tau\right)}  \right]^T \in \mathbb{C}^{\tau \times 1}$ denotes the AWGN vector.
Based on (\ref{on_sigs}), the ML estimation of all the relevant unknown channel parameters\footnote{We assume that $\boldsymbol{\bar{b}}_1$ can be perfectly estimated off-line for simplicity. Note that its estimation error can be reduced to a very low value with sufficient training time during the off-line estimation.
} is given by (with irrelevant terms omitted)
\begin{align}\label{mle_c}
&\{\hat{\bar{a}}^{[n]},\hat{\vartheta}^{[n]},\hat{\psi}^{[n]},\hat{d}^{[n]}_{\mathrm{C}}\} 
= \arg \min _{\bar{a}^{[n]},\vartheta^{[n]},\psi^{[n]} , d^{[n]}_{\mathrm{C}}}  \left\|\boldsymbol{y}^{[n]} -\bar{a}^{[n]} \boldsymbol{V} \operatorname{diag}\left(\boldsymbol{\bar{b}}_1 \right)
\boldsymbol{u}\left({\vartheta^{[n]}},\psi^{[n]}\right) - d^{[n]}_{\mathrm{C}} \boldsymbol{1}_{\tau }\right\|^2.
\end{align}
Since there are four unknown parameters in (\ref{mle_c}), $\tau \geq \operatorname{rank}\left(\boldsymbol{V}\right) \geq 4$ is generally required for the ML estimation problem in (\ref{mle_c}) to ensure its feasibility.
However, the estimation in (\ref{mle_c}) incurs prohibitively high computational complexity due to the joint search over $\{\bar{a}^{[n]},\vartheta^{[n]},\psi^{[n]} , d^{[n]}_{\mathrm{C}}\}$. It is noted that only $\{\vartheta^{[n]}, \psi^{[n]}\}$ are needed for the passive beamforming design in (\ref{online_pbf}). To tackle the above-mentioned issues, the estimation of $\{\vartheta^{[n]}, \psi^{[n]}\}$ is decoupled as follows. 
First, with fixed $\boldsymbol{V}$ and given $\{\bar{a}^{[n]},\vartheta^{[n]},\psi^{[n]}\}$, the optimal estimate of $d^{[n]}_{\mathrm{C}}$ to minimize the objective function in (\ref{mle_c}) is given by
\begin{equation}\label{mle_c1}
\hat{d}^{[n]}_{\mathrm{C}} = \frac{\boldsymbol{1}_{\tau}^T \left(\boldsymbol{y}^{[n]} -\bar{a}^{[n]} \boldsymbol{V} \operatorname{diag}\left(\boldsymbol{\bar{b}}_1 \right)
\boldsymbol{u}\left({\vartheta^{[n]}},\psi^{[n]}\right)\right)}{\tau}.
\end{equation}
Substituting (\ref{mle_c1}) into (\ref{mle_c}), the estimates of $\{\bar{a}^{[n]},\vartheta^{[n]},\psi^{[n]}\}$ are given by
\begin{equation}\label{mle_c2}
\{\hat{\bar{a}}^{[n]},\hat{\vartheta}^{[n]},\hat{\psi}^{[n]}\} = \arg \min _{a^{[n]},\vartheta^{[n]},\psi^{[n]}} \left\|\bar{\boldsymbol{y}}^{[n]} -\bar{a}^{[n]} \boldsymbol{\eta}\left({\vartheta},\psi \right)\right\|^2,
\end{equation}
where we define $\bar{\boldsymbol{y}}^{[n]} = \left(\boldsymbol{\mathrm{I}}_{\tau} -\frac{ \boldsymbol{1}_{\tau } \boldsymbol{1}_{\tau }^T}{\tau}\right) \boldsymbol{y}^{[n]}$ and $\boldsymbol{\eta}\left({\vartheta},\psi\right) = \left(\boldsymbol{\mathrm{I}}_{\tau} -\frac{ \boldsymbol{1}_{\tau } \boldsymbol{1}_{\tau }^T}{\tau}\right)\boldsymbol{V} \operatorname{diag}\left(\boldsymbol{\bar{b}}_1 \right) \boldsymbol{u}\left({\vartheta},\psi\right)$ for notational convenience. For given $\{\vartheta^{[n]},\psi^{[n]}\}$, the optimal estimate of $\bar{a}^{[n]}$ to minimize the objective function in (\ref{mle_c2}) is given by
\begin{equation}\label{mle_c3}
\hat{\bar{a}}^{[n]} = \frac{\boldsymbol{\eta}^H\left({\vartheta}^{[n]},\psi^{[n]}\right)\bar{\boldsymbol{y}}^{[n]}}{\left\|\boldsymbol{\eta}\left({\vartheta}^{[n]},\psi^{[n]}\right) \right\|^2}
.
\end{equation}
Substituting (\ref{mle_c3}) into (\ref{mle_c2}), the ML estimates of $\{\vartheta^{[n]},\psi^{[n]}\}$ are given by
\begin{equation}\label{mle_c4}
\{ \hat{\vartheta}^{[n]},\hat{\psi}^{[n]}\} = \arg \max _{\vartheta^{[n]},\psi^{[n]}} \frac{\left|\boldsymbol{\eta}^H\left({\vartheta}^{[n]},\psi^{[n]}\right)\bar{\boldsymbol{y}}^{[n]}\right|^2}{\left\|\boldsymbol{\eta}\left({\vartheta}^{[n]},\psi^{[n]}\right) \right\|^2}
.
\end{equation}
It is observed that the scaling factor ambiguity in $\boldsymbol{\bar{b}}_1$ will not affect the estimation of $\{\vartheta^{[n]}, \psi^{[n]}\}$. 
Note that the problem in (\ref{mle_c4}) is still a non-convex optimization problem as its objective function is non-concave with respect to $\vartheta^{[n]}$ and $\psi^{[n]}$, which is thus difficult to be solved optimally.
The two-step gradient based search proposed in \cite{vehicle-side-irs} can be applied to obtain a locally optimal solution for this problem.
The main computation burden of solving (\ref{mle_c4}) via the two-step gradient based search proposed in \cite{vehicle-side-irs} lies in the initial search and the gradient search for $\{\vartheta^{[n]},\psi^{[n]}\}$, whose complexity orders are given by $\mathcal{O}\left( M \tau A_1 \right)$ and $\mathcal{O}\left( M \tau I_1 \right)$, respectively, with $A_1$ and $I_1$ respectively denoting the grid size for the initial search and number of iterations required for convergence. As such, the total computational complexity order is given by $\mathcal{O}\left( M \tau \left(A_1 + I_1 \right) \right)$.
When the estimated $\{ \hat{\vartheta}^{[n]},\hat{\psi}^{[n]}\}$ reaches a pre-defined coverage region, C\textsubscript{1} sets $n = 0$ and starts the channel/beam prediction, as elaborated next.

\subsubsection{Passive Beam Prediction for $n > 0$ (when IRS\textsubscript{1} is serving the user).}
\begin{figure}
\centering
\includegraphics[width=0.65\textwidth]{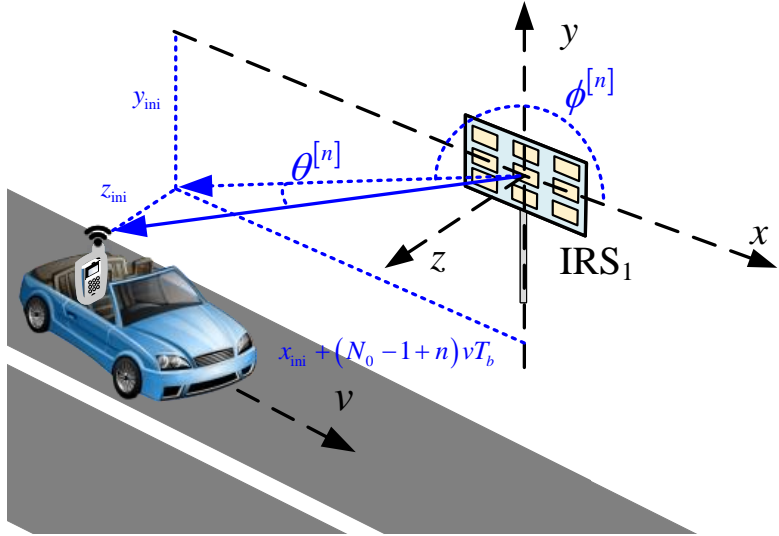}
\caption{Illustration of the geometric location of the high-mobility user.}
\label{geometric}
\end{figure}
With $\{\hat{\vartheta}^{[n]},\hat{\psi}^{[n]}\}$ obtained for $n \leq 0$, we aim to predict $\{\hat{\vartheta}^{[n]},\hat{\psi}^{[n]}\}$ for $n > 0$ (i.e., $n\in \mathcal{N}$).
Let $N_0$ denote the number of blocks when C\textsubscript{1} has estimated $\{\vartheta^{[n]},\psi^{[n]}\}$ based on (\ref{mle_c4}) for $n \leq 0$ and $\mathcal{N}_0 \triangleq \{1-N_0, \ldots,-1,0\}$ denote the index set.
Denote the initial relative location of the user with respect to IRS\textsubscript{1} as $[x_\mathrm{ini},y_\mathrm{ini},z_\mathrm{ini}]$ in block $1-N_0$. 
Hence, the relative location of the user in block $n$ is given by $[x^{[n]},y_\mathrm{ini},z_\mathrm{ini}]$ with $x^{[n]} =  x_\mathrm{ini}+(n + N_0-1)v T_b $, as illustrated in Fig.~\ref{geometric}. 
Based on the mapping from the spherical coordinate to the 3D Cartesian coordinate, it can be verified that $\vartheta^{[n]}$ and $\psi^{[n]}$ are functions of $\{x_\mathrm{ini},y_\mathrm{ini},z_\mathrm{ini},v\}$ given by
\begin{align}\label{varthe_fun}
\vartheta^{[n]} &= \frac{\lambda x^{[n]}}{2d_\mathrm{I}\sqrt{\left(x^{[n]}\right)^2 +y_\mathrm{ini}^2 + z_\mathrm{ini}^2}},
\psi^{[n]} &= \frac{\lambda y_\mathrm{ini}}{2d_\mathrm{I}\sqrt{\left(x^{[n]}\right)^2 +y_\mathrm{ini}^2 + z_\mathrm{ini}^2}}, \quad \forall n.
\end{align} 
Based on the estimates of $\{\vartheta^{[n]},\psi^{[n]}\}_{n\in \mathcal{N}_0}$, it is possible to further resolve the unknown parameters $\{x_\mathrm{ini},y_\mathrm{ini},z_\mathrm{ini},v\}$ in (\ref{varthe_fun}), which can be used for predicting $\{\vartheta^{[n]},\psi^{[n]}\}_{n\in \mathcal{N}}$.
However, the joint search over $\{x_\mathrm{ini},y_\mathrm{ini},z_\mathrm{ini},v\}$ requires prohibitive computational complexity. To tackle this difficulty, we propose to only estimate the free parameters in $\{x_\mathrm{ini},y_\mathrm{ini},z_\mathrm{ini},v\}$ by exploiting the geometric properties of the relevant parameters.
As shown in Fig.~\ref{geometric}, based on the geometric definitions of the elevation and azimuth AoAs (i.e., $\{{\theta}^{[n]}, {\phi}^{[n]}\}$ in (\ref{user-irs})) from the user to IRS\textsubscript{1}, we have 
\begin{equation}\label{trigo_1}
\frac{y_\mathrm{ini}}{x^{[n]}} = \tan {\phi}^{[n]},\quad \frac{z_\mathrm{ini}}{\sqrt{\left( x^{[n]}\right)^2 + y_\mathrm{ini}^2}} = \tan {\theta}^{[n]}, \quad \forall n.
\end{equation} 
According to (\ref{trigo_1}), by eliminating $y_\mathrm{ini}$, we have
\begin{align}\label{elim_y_sing}
\frac{z_\mathrm{ini}}{ x^{[n]}} &= \frac{z_\mathrm{ini}}{  x_\mathrm{ini}+(n + N_0-1)v T_b} = \underbrace{\tan {\theta}^{[n]}  \sqrt{1+ \tan^2 {\phi}^{[n]}}}_{r^{[n]}},\quad \forall n.
\end{align} 
By stacking $N_0$ equations in (\ref{elim_y_sing}) over $n \in \mathcal{N}_0$, we have
\begin{equation}\label{elim_y_vec}
z_\mathrm{ini} \boldsymbol{\epsilon}\left(x_\mathrm{ini},v\right) = \boldsymbol{r},
\end{equation} 
where $\boldsymbol{\epsilon}\left(x_\mathrm{ini},v\right) = \left[\frac{1}{x_\mathrm{ini}},\ldots, \frac{1}{x_\mathrm{ini}+(N_0-1)v T_b}\right]^T \in \mathbb{R}^{N_0 \times 1}$ and $\boldsymbol{r} = \left[ r^{[1-N_0]},\ldots,r^{[0]}\right]^T \in \mathbb{R}^{N_0 \times 1}$
Note that given any $\{\vartheta^{[n]},\psi^{[n]}\}$, based on the relationship among trigonometric functions, $\tan {\theta}^{[n]}$ and $\tan {\phi}^{[n]}$ can be respectively calculated by 
\begin{align}\label{tan_the}
\tan {\theta}^{[n]} &= \sqrt{\frac{1}{\cos^2{\theta}^{[n]}}- 1}\stackrel{(a)}{=} \sqrt{ \frac{\lambda^2/\left(4 d_\mathrm{I}^2\right)}{\left({\psi}^{[n]}\right)^2 +  \left({\vartheta}^{[n]}\right)^2} -1 },\quad \forall n,
\end{align} 
\begin{equation}\label{tan_phi}
\tan {\phi}^{[n]} = \frac{\psi^{[n]}}{\vartheta^{[n]}}, \quad \forall n,
\end{equation} 
where $(a)$ holds due to the fact that $\left({\psi}^{[n]}\right)^2 +  \left({\vartheta}^{[n]}\right)^2 = \left(\frac{2d_\mathrm{I}}{\lambda}\right)^2 \cos^2{\theta}^{[n]}$.
Substituting the estimated $\{\hat{\vartheta}^{[n]},\hat{\psi}^{[n]}\}_{n \in \mathcal{N}_0}$ into (\ref{tan_the}) and (\ref{tan_phi}), the estimate of $r^{[n]}$ is given by
\begin{equation}\label{ssss4}
\hat{r}^{[n]} =  \sqrt{\left( \frac{\lambda^2/\left(4d_\mathrm{I}^2\right)}{\left(\hat{\psi}^{[n]}\right)^2 +  \left(\hat{\vartheta}^{[n]}\right)^2} -1 \right) \left(1+ \left(\frac{\hat{\psi}^{[n]}}{\hat{\vartheta}^{[n]}}\right)^2\right)},  n \in \mathcal{N}_0.
\end{equation} 
Based on (\ref{elim_y_vec}) and (\ref{ssss4}), by stacking $N_0$ estimated scalars $\{\hat{r}^{[n]}\}_{n \in \mathcal{N}_0}$ into the vector $\boldsymbol{\hat{r}} = \left[\hat{r}^{[1-N_0]},\ldots,\hat{r}^{[0]}\right]^T \in \mathbb{R}^{N_0 \times1}$, the estimation problem of the relevant parameters can be formulated as 
\begin{equation}\label{ls_po}
\{\hat{z}_\mathrm{ini},\hat{x}_\mathrm{ini},\hat{v}\} = \arg \min _{z_\mathrm{ini},x_\mathrm{ini},v} 
\left\|\boldsymbol{\hat{r}} -z_\mathrm{ini} \boldsymbol{\epsilon}\left(x_\mathrm{ini},v \right)\right\|^2.
\end{equation}
Since there are three unknown parameters in (\ref{ls_po}), $N_0  \geq 3$ is generally required for the estimation problem in (\ref{ls_po}) to ensure its feasibility.
Given $\{x_\mathrm{ini},v\}$, the optimal value of $z_\mathrm{ini}$ to minimize the objective function in (\ref{ls_po}) is given by
\begin{equation}\label{ls_z}
\hat{z}_\mathrm{ini} = \frac{\boldsymbol{\epsilon}^T\left(x_\mathrm{ini},v\right)\boldsymbol{\hat{r}}}{\left\|\boldsymbol{\epsilon}\left(x_\mathrm{ini},v\right) \right\|^2}.
\end{equation}
Substituting (\ref{ls_z}) into (\ref{ls_po}), the estimates of $\{x_\mathrm{ini},v\}$ are given by
\begin{equation}\label{ls_c4}
\{ \hat{x}_\mathrm{ini},\hat{v}\} = \arg \max _{x_\mathrm{ini},v} \frac{\left|\boldsymbol{\epsilon}^T\left(x_\mathrm{ini},v\right) \boldsymbol{\hat{r}}\right|^2}{\left\|\boldsymbol{\epsilon}\left(x_\mathrm{ini},v\right) \right\|^2}.
\end{equation}
It can be verified that the problem in (\ref{ls_c4}) is still a non-convex optimization problem as its objective function is non-concave with respect to $x_\mathrm{ini}$ and $v$, which is thus difficult to be solved optimally. The two-step gradient based search proposed in \cite{vehicle-side-irs} can be similarly applied to obtain a locally optimal solution for this problem.
The main computation burden of solving (\ref{ls_c4}) via the two-step gradient based search proposed in \cite{vehicle-side-irs} lies in the initial search and the gradient search for $\{x_\mathrm{ini},v\}$, whose complexity orders are $\mathcal{O}\left( N_0  A_2 \right)$ and $\mathcal{O}\left( N_0  I_2 \right)$, respectively, with $A_2$ and $I_2$ respectively denoting the grid size for the initial search and number of iterations required for convergence. As such, the total computational complexity order is given by $\mathcal{O}\left( N_0  \left(A_2 + I_2 \right) \right)$. It is worth noting that solving (\ref{ls_c4}) only needs to be performed once for each IRS serving the user.
With $\{ \hat{x}_\mathrm{ini},\hat{v}\}$ obtained in (\ref{ls_c4}), the estimate of $z_\mathrm{ini}$ can be obtained according to (\ref{ls_z}). Moreover, based on (\ref{trigo_1}) and (\ref{tan_phi}), the estimate of $y_\mathrm{ini}$ is obtained as
\begin{equation}\label{y_est}
\hat{y}_\mathrm{ini} = \frac{1}{N_0} \sum^0_{n = 1-N_0}  \frac{\hat{\psi}^{[n]}\left(\hat{x}_\mathrm{ini}+(n + N_0-1)\hat{v} T_b\right)}{\hat{\vartheta}^{[n]}}.
\end{equation} 
Then, the predicted $\{\hat{\vartheta}^{[n]},\hat{\psi}^{[n]}\}_{n \in \mathcal{N}}$ can be calculated by substituting $\{\hat{x}_\mathrm{ini},\hat{y}_\mathrm{ini},\hat{z}_\mathrm{ini},\hat{v}\}$ into (\ref{varthe_fun}), based on which C\textsubscript{1} dynamically sets the passive beamforming vector of IRS\textsubscript{1} according to (\ref{online_pbf}).

\section{Simulation Results}
\begin{figure}
\centering
\includegraphics[width=0.7\textwidth]{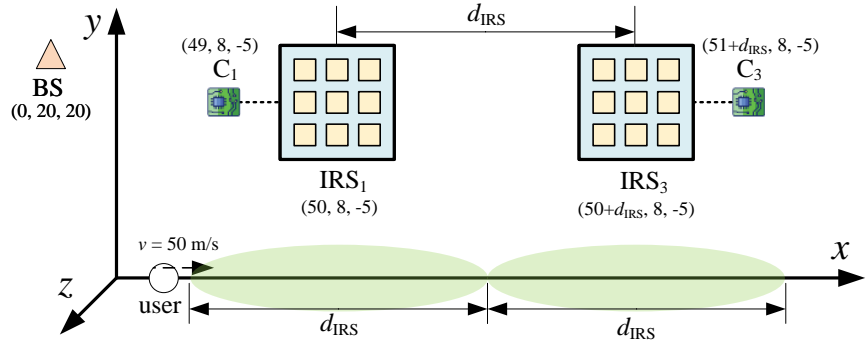}
\caption{Initial setup for simulations.}
\label{loca}
\end{figure}
In this section, we present simulation results to evaluate the performance of our proposed roadside IRS-aided high-mobility communication system.
We set the carrier frequency as $f_c = 5.9$~GHz, as specified in the 3GPP standard in \cite{cv2x2} for cellular V2X (C-V2X) applications, with the signal bandwidth of 1~MHz. 
The vehicle speed is set as $v=50$~m/s (if not specified otherwise), which results in a Doppler frequency with the maximum value of $f_{max} = v f_c/c  \approx 1 $~KHz, where $c = 3 \times 10^8$~m/s denotes the speed of light.
The duration of each block is set as $T_b = 1/\left(10 f_{max}\right) \approx 0.1$~millisecond (ms), during which all the channels are assumed to remain approximately constant.
We set the half-wavelength spacing for the adjacent BS antennas and IRS reflecting elements.
The initial setup of the system is shown in Fig.~\ref{loca}, where $d_{\mathrm{IRS}}$ denotes the inter-IRS/coverage distance. 
For the purpose of exposition, we consider a typical communication period during which the user is passing by IRS\textsubscript{1}, with $N =  \lceil\frac{d_{\mathrm{IRS}}}{v T_b}\rceil$.
The general clustered delay line model in the 3GPP TR 38.901 Release 16 \cite[Table 7.7.1-4]{3gpp_pg} (assuming the frequency-flat fading over the sub-bandwidth of 1~MHz) is adopted to generate the IRS\textsubscript{1}\textrightarrow BS channel $\boldsymbol{G}$, where we set the number of paths $L=3$ and randomly generate the AoA and AoDs of each path within their defined range.
The path loss exponents of the IRS\textsubscript{1}\textrightarrow BS and user\textrightarrow BS channels are set as 2.1 and 2.5, respectively, while the path loss exponents of all the LoS channels are set as 2.
The channel power gain at the reference distance of 1~m is set as $\beta_0 = -30$~dB for each link. 
We assume that the (direct) user\textrightarrow BS channel $\boldsymbol{d}^{[n]}$ and (direct) user\textrightarrow C\textsubscript{1} channel $d_\mathrm{C}^{[n]}$, $n\in \mathcal{N}$, follow the Rayleigh fading channel model with the time correlation modeled by the Jakes' spectrum\cite{jakes}. 
Moreover, the C\textsubscript{1}\textrightarrow IRS\textsubscript{1} channel $\boldsymbol{b}_1$ is assumed to be modeled by the near-field LoS channel. 
Let $P_t$ denote the transmit power at the user and the noise power at the BS is set as $\sigma_\mathrm{B}^2 = -70$~dBm. Accordingly, the normalized noise power at the BS is given by $\sigma^2 = \sigma_\mathrm{B}^2/P_t$.
\subsection{Performance of Proposed Channel Estimation Scheme}
First, we evaluate the performance of our proposed channel estimation scheme. We consider the average achievable rate with the training overhead taken into account as the performance metric in the rest of this paper (if not specified otherwise), where the performance gap (due to the practical modulation and coding scheme) to the capacity is set as 9~dB \cite{vehicle-side-irs}.
We take the case with the perfect channel angle/phase information on $\{\vartheta^{[n]},\psi^{[n]}\}$ as the performance upper bound, where we have $\{\hat{\vartheta}^{[n]},\hat{\psi}^{[n]}\}=\{\vartheta^{[n]},\psi^{[n]}\}$ for $n\in\mathcal{N}$.
\begin{figure}
\centering
\includegraphics[width=0.6\textwidth]{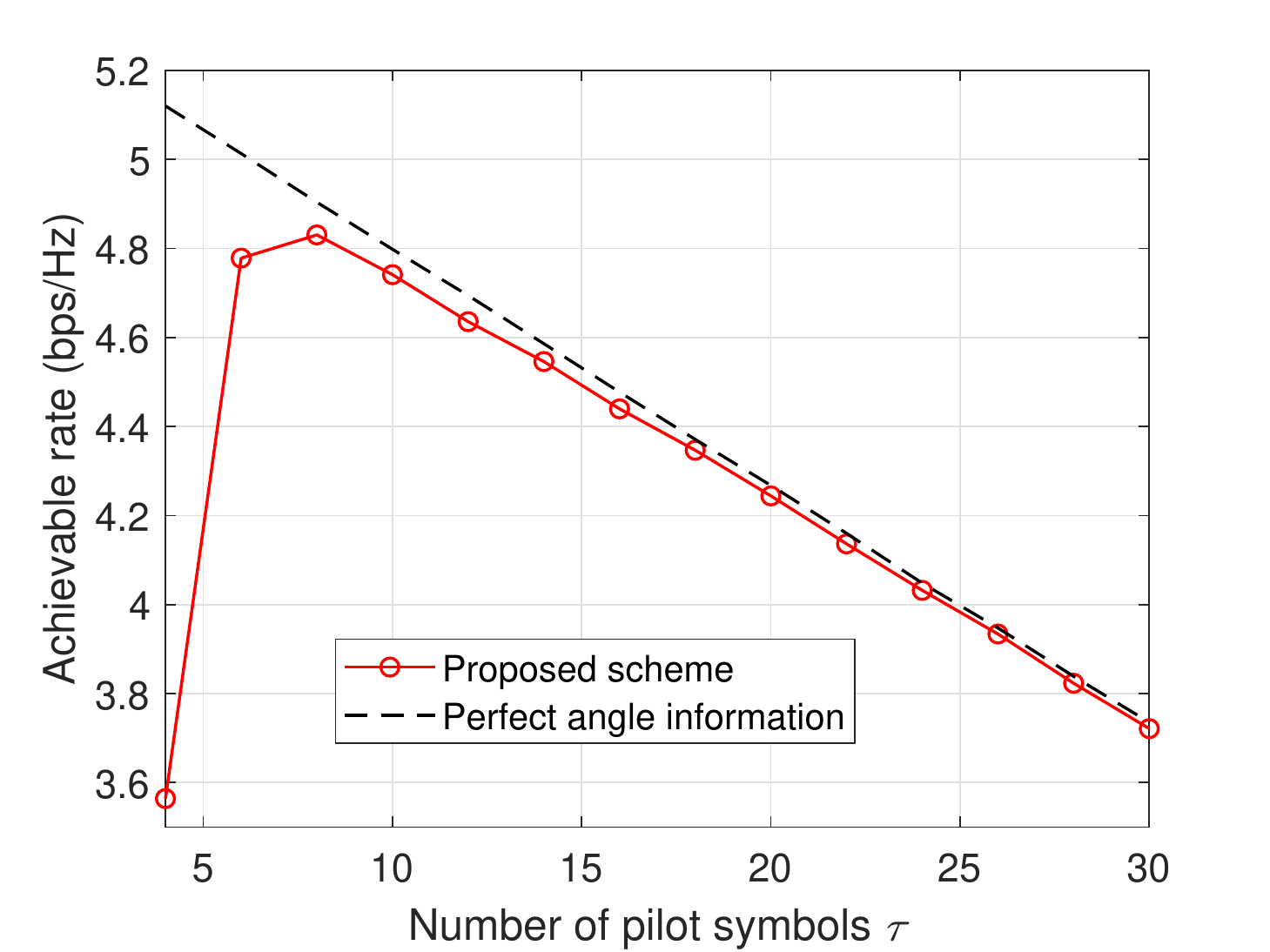}
\caption{Achievable rate versus the number of pilot symbols in each block, $\tau$.}
\label{rate_tau}
\end{figure}
In Fig.~\ref{rate_tau}, we show the achievable rate versus the number of pilot symbols in each block, $\tau$, with $P_t = 12$~dBm, $M = 200$, $d_{\mathrm{IRS}} = 4$~m, $N_\mathrm{B} = 16$, and $N_0 = 30$. 
It is observed that there exists a trade-off in the time/symbol allocation between the uplink training and data transmission in each block for maximizing the achievable rate, which can be explained as follows.
With too little training, the estimates of $\{\vartheta^{[n]},\psi^{[n]}\}_{n\in \mathcal{N}_0}$ are not accurate enough for predicting $\{\vartheta^{[n]},\psi^{[n]}\}_{n\in \mathcal{N}}$, which degrades the passive beamforming gain, while too much training results in less time for reaping the high passive beamforming gain for data transmission, thus both degrading the throughput of the system.
\begin{figure}
\centering
\includegraphics[width=0.6\textwidth]{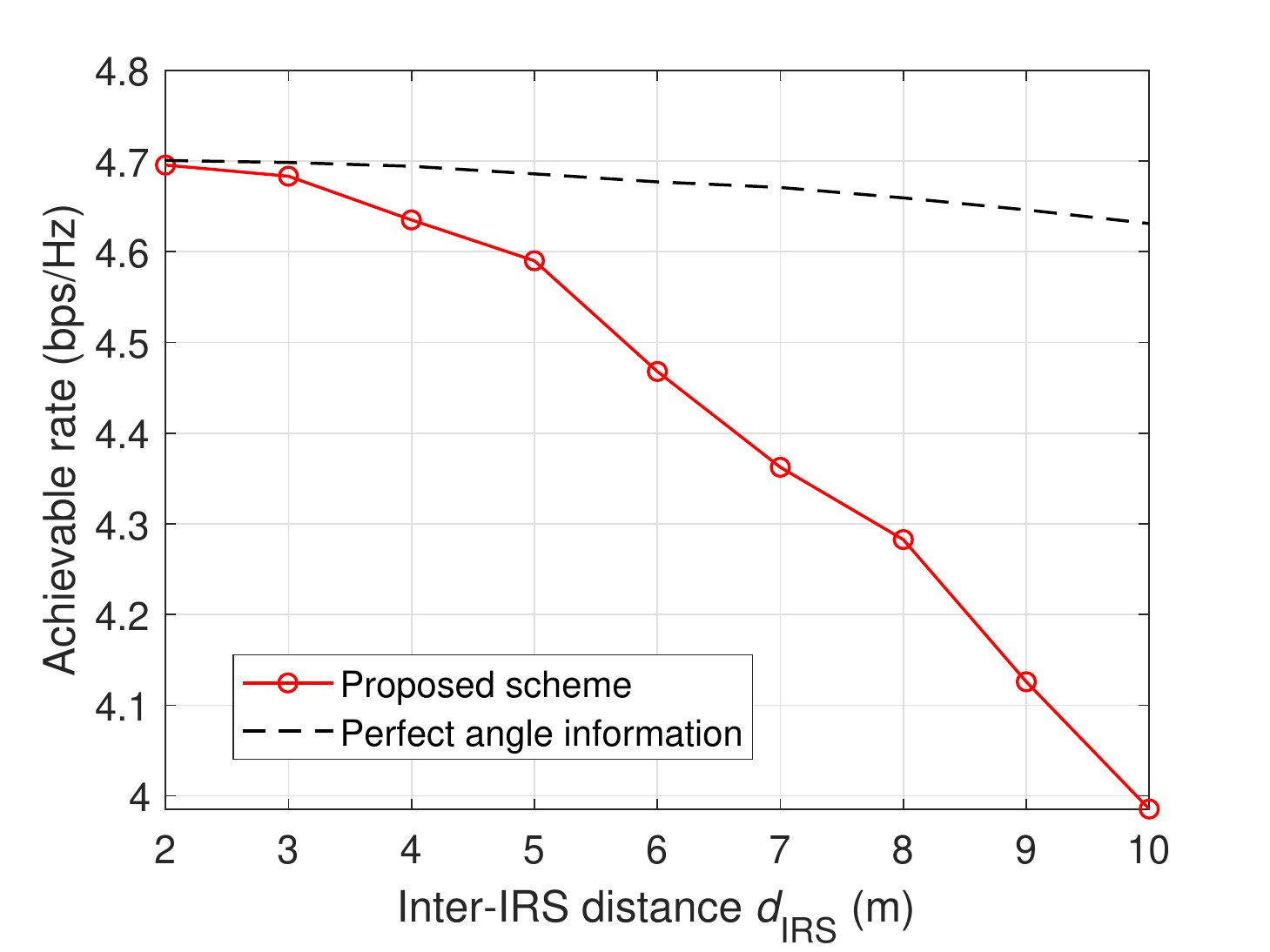}
\caption{Achievable rate versus the inter-IRS distance, $d_{\mathrm{IRS}}$.}
\label{rate_d}
\end{figure}

In Fig.~\ref{rate_d}, we show the achievable rate versus the inter-IRS distance $d_{\mathrm{IRS}}$, with $P_t = 12$~dBm, $M = 200$, $\tau = 10$, $N_\mathrm{B} = 16$, and $N_0 = 30$. It is observed that the achievable rate of the proposed scheme decreases with increasing $d_{\mathrm{IRS}}$. Moreover, the performance gap between the proposed scheme and the performance upper bound also increases with increasing $d_{\mathrm{IRS}}$. This is due to the fact that with the fixed vehicle speed $v$, the serving period of each IRS becomes longer (i.e., $N$ is larger) with larger $d_{\mathrm{IRS}}$. 
Due to the accumulative prediction errors, the predicted $\{\hat{\vartheta}^{[n]},\hat{\psi}^{[n]}\}_{n\in \mathcal{N}}$ may deviate from the actual $\{\vartheta^{[n]},\psi^{[n]}\}_{n\in \mathcal{N}}$, which thus degrades the passive beamforming gain and also the achievable rate.
\subsection{IRS-Enhanced Channel}
Next, we show the IRS-enhanced channel from the moving user to the BS over time using our proposed scheme\footnote{For ease of illustration, we consider only two IRSs (IRS 1 and IRS 3) as shown in Fig.~\ref{loca}.}, as compared to the case without IRS. 
Let $\gamma^{[n]}= \left\|\boldsymbol{Q}^{[n]} \boldsymbol{\nu}^{[n]}  + \boldsymbol{d}^{[n]}\right\|^2$ denote the effective channel power gain, which is taken as the performance metric.
\begin{figure}
\centering
\includegraphics[width=0.6\textwidth]{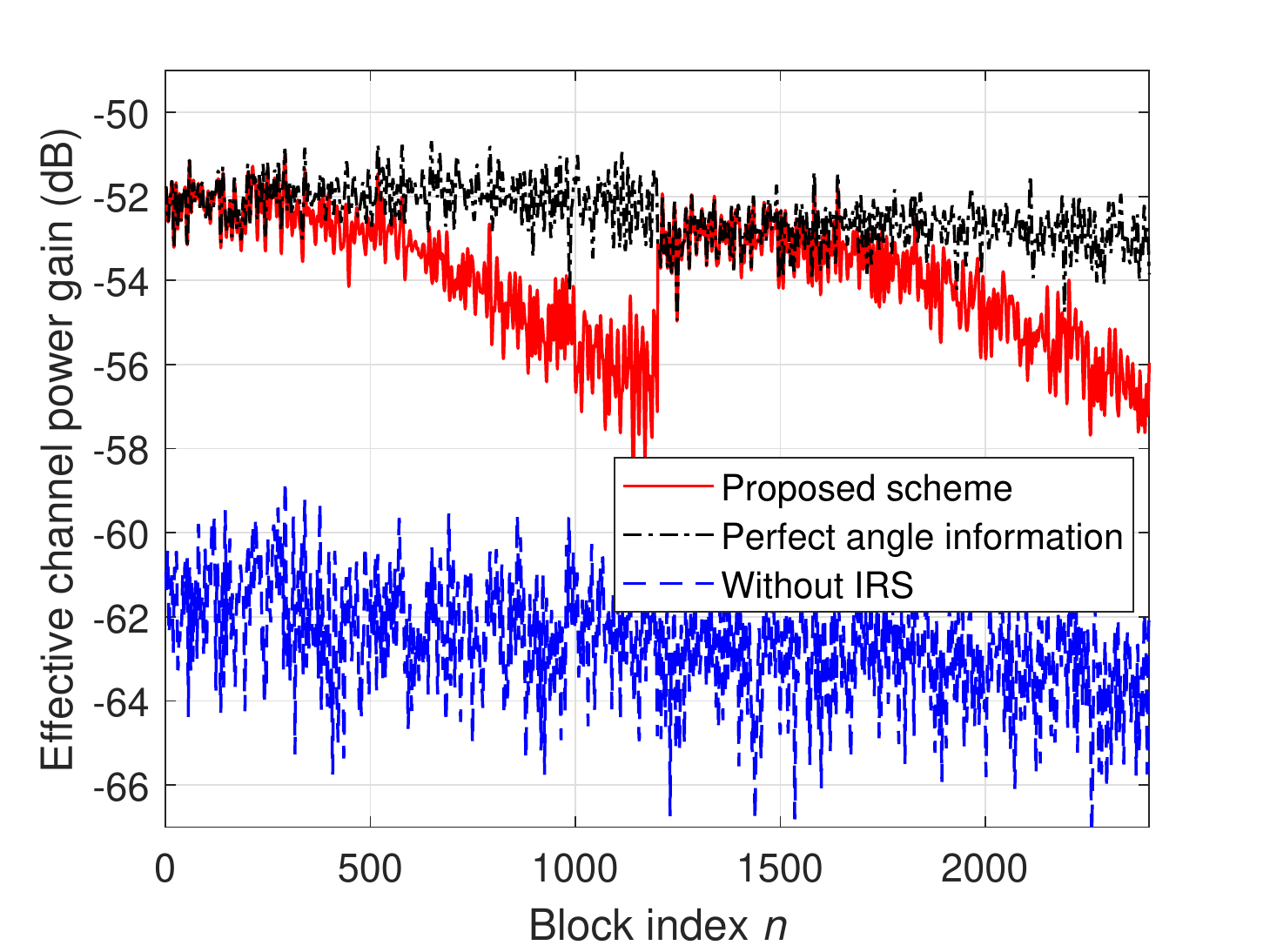}
\caption{Effective channel power gain over time.}
\label{snr_time}
\end{figure}
In Fig.~\ref{snr_time}, we show one realization of $\gamma^{[n]}$, $n \in \mathcal{N}$, over time, with $P_t = 12$~dBm, $M = 400$, $d_{\mathrm{IRS}} = 6$~m, $\tau = 10$, $N_\mathrm{B} = 16$, and $N_0 = 30$. 
It is observed that the proposed scheme not only achieves a higher average channel gain but also leads to less channel gain fluctuation (i.e., less fading), as compared to the benchmark scheme without IRS.
Moreover, it is observed that the effective channel gain of the proposed scheme decreases over time, which is due to the channel estimation/prediction error that accumulates over time and thus causes more deviations of the IRS passive beamforming direction from the optimal one assuming the perfect channel information.

\subsection{Performance Comparison with Benchmark Schemes}
In this subsection, we evaluate the achievable rate performance of our proposed channel estimation scheme by comparing it with the following benchmark schemes. 
\begin{enumerate}
    \item \textbf{Upper bound:}
    We consider the performance upper bound by solving (P1) based on the perfect CSI of both $\boldsymbol{Q}^{[n]}$ and $\boldsymbol{d}^{[n]}$, $n \in \mathcal{N}$, for comparison.
    
    \item \textbf{Channel Estimation Scheme in \cite{liu}:} 
    $\boldsymbol{Q}^{[1]}$ and $\boldsymbol{d}^{[1]}$ are firstly estimated at the BS with $M+1$ pilot symbols in the first block; with $\boldsymbol{Q}^{[1]}$ exploited as the reference CSI, the BS then estimates $\boldsymbol{Q}^{[n]}$ and $\boldsymbol{d}^{[n]}$ for $n > 1$ with the minimum training overhead of $1+ \lceil \frac{M}{N_\mathrm{B}} \rceil$. Based on the estimated CSI, IRS sets the passive beamforming $\boldsymbol{\nu}^{[n]}$ by solving (P2) for data transmission for $n\in \mathcal{N}$. Note that this scheme requires the signal feedback from the BS to the IRSC for setting IRS's reflection with the feedback overhead in the order of $\mathcal{O}\left(M\right)$, which may cause significant delay and outdated IRS passive beamforming direction. However, we ignore such delay in our comparison in favor of the scheme in \cite{liu}.
    
    \item \textbf{Channel Estimation Scheme in \cite{dll}:} 
    Assuming that the IRS\textrightarrow BS channel $\boldsymbol{G}$ can be perfectly resolved off-line, the BS estimates $\boldsymbol{Q}^{[n]}$ and $\boldsymbol{d}^{[n]}$ for each block $n$ with the minimum training overhead of $1+ \lceil \frac{M}{N_\mathrm{B}} \rceil$.
    Based on the estimated CSI, IRS sets the passive beamforming $\boldsymbol{\nu}^{[n]}$ by solving (P2) for data transmission with $n\in \mathcal{N}$. Note that this scheme also requires the signal feedback from the BS to the IRSC for setting IRS's reflection with the feedback overhead in the order of $\mathcal{O}\left(M\right)$, which is ignored in our comparison in favor of the scheme in \cite{dll}.
    
\end{enumerate}
\begin{figure}
\centering
\includegraphics[width=0.6\textwidth]{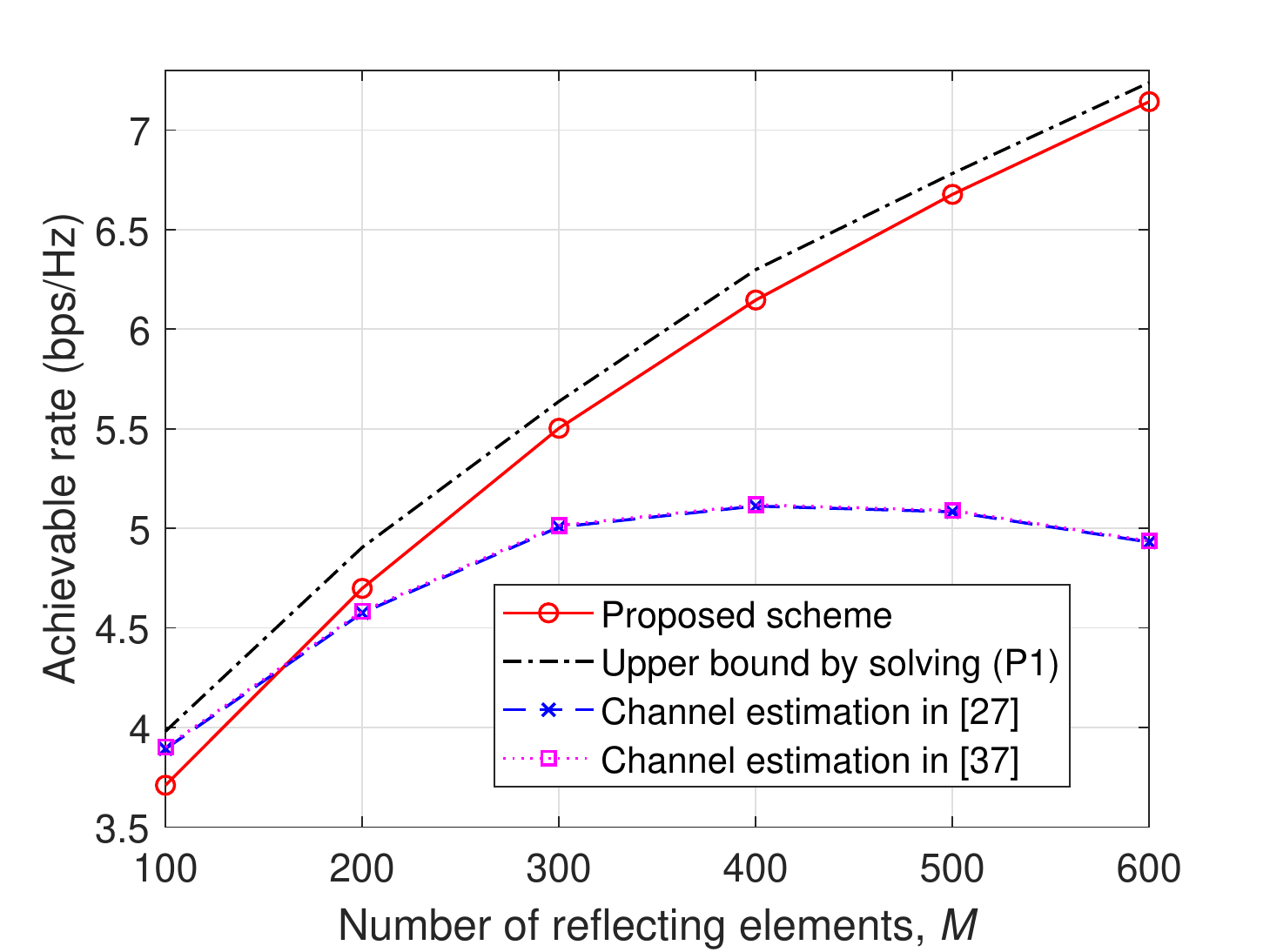}
\caption{Achievable rate versus the number of reflecting elements, $M$.}
\label{rate_m}
\end{figure}

In Fig.~\ref{rate_m}, we show the achievable rate versus the number of reflecting elements $M$, with $P_t = 12$~dBm, $d_{\mathrm{IRS}} = 2$~m, $\tau = 12$, $N_\mathrm{B} = 16$, and $N_0 = 30$.
It is observed that the achievable rate of the proposed scheme increases with $M$. This is due to the fact that the proposed scheme efficiently estimates/predicts $\{\vartheta^{[n]},\psi^{[n]}\}$ without the need of increasing the training overhead and also achieves a higher passive beamforming gain as $M$ increases. 
In contrast, the achievable rate of the schemes in \cite{liu} and \cite{dll} first increases and then substantially decreases with $M$, which is due to the increasing training overhead with $M$ required for the cascaded channel estimation.
It is also observed that the channel estimation scheme in \cite{dll} achieves slightly better rate performance as compared to the scheme in \cite{liu}, which is owing to the off-line estimation of the BS-IRS channel assumed in \cite{dll} for reducing the training overhead in the first block.
Moreover, it is observed that the proposed scheme (which maximizes the IRS-reflected channel gain only) incurs some performance loss, as compared to the upper bound that aligns both the IRS-reflected and non-IRS-reflected channels assuming the perfect CSI. 
Nevertheless, such performance gap of the proposed scheme becomes smaller with a larger $M$. This is due to the fact that with a large $M$, the performance gain by aligning the IRS-reflected channel with the non-IRS-reflected channel becomes marginal, as the power of the former becomes dominant over that of the latter and also it becomes more difficult to align these channels with larger values of $M$.

\begin{figure}
\centering
\includegraphics[width=0.6\textwidth]{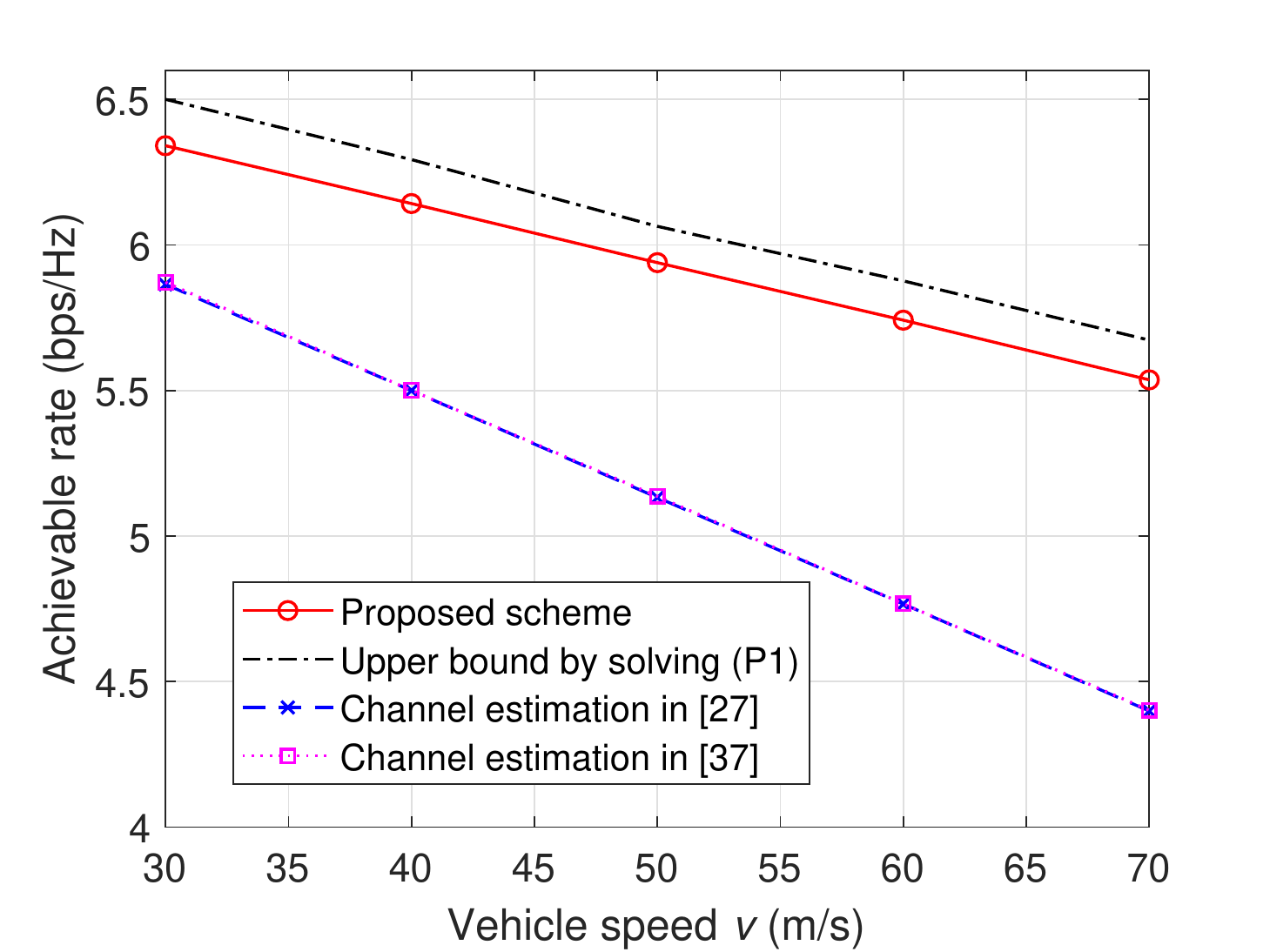}
\caption{Achievable rate versus the vehicle speed, $v$.}
\label{rate_v}
\end{figure}
Next, we show the achievable rate versus the vehicle speed $v$ in Fig.~\ref{rate_v}, with $P_t = 12$~dBm, $M = 400$, $d_{\mathrm{IRS}} = 4$~m, $\tau = 15$, $N_\mathrm{B} = 16$, and $N_0 = 30$.
It is observed that the achievable rates of both the proposed scheme and the benchmark schemes in \cite{liu} and in \cite{dll} decrease with the increasing vehicle speed $v$. This is due to the fact that with the higher user mobility, the channel coherence time becomes smaller, thus making the time left for data transmission become shorter given the same training overhead. Nevertheless, the proposed scheme exhibits a lower decreasing rate with $v$, which is due to its lower training overhead as compared to that of the benchmark schemes in \cite{liu} and \cite{dll}. 

\begin{figure}
\centering
\includegraphics[width=0.6\textwidth]{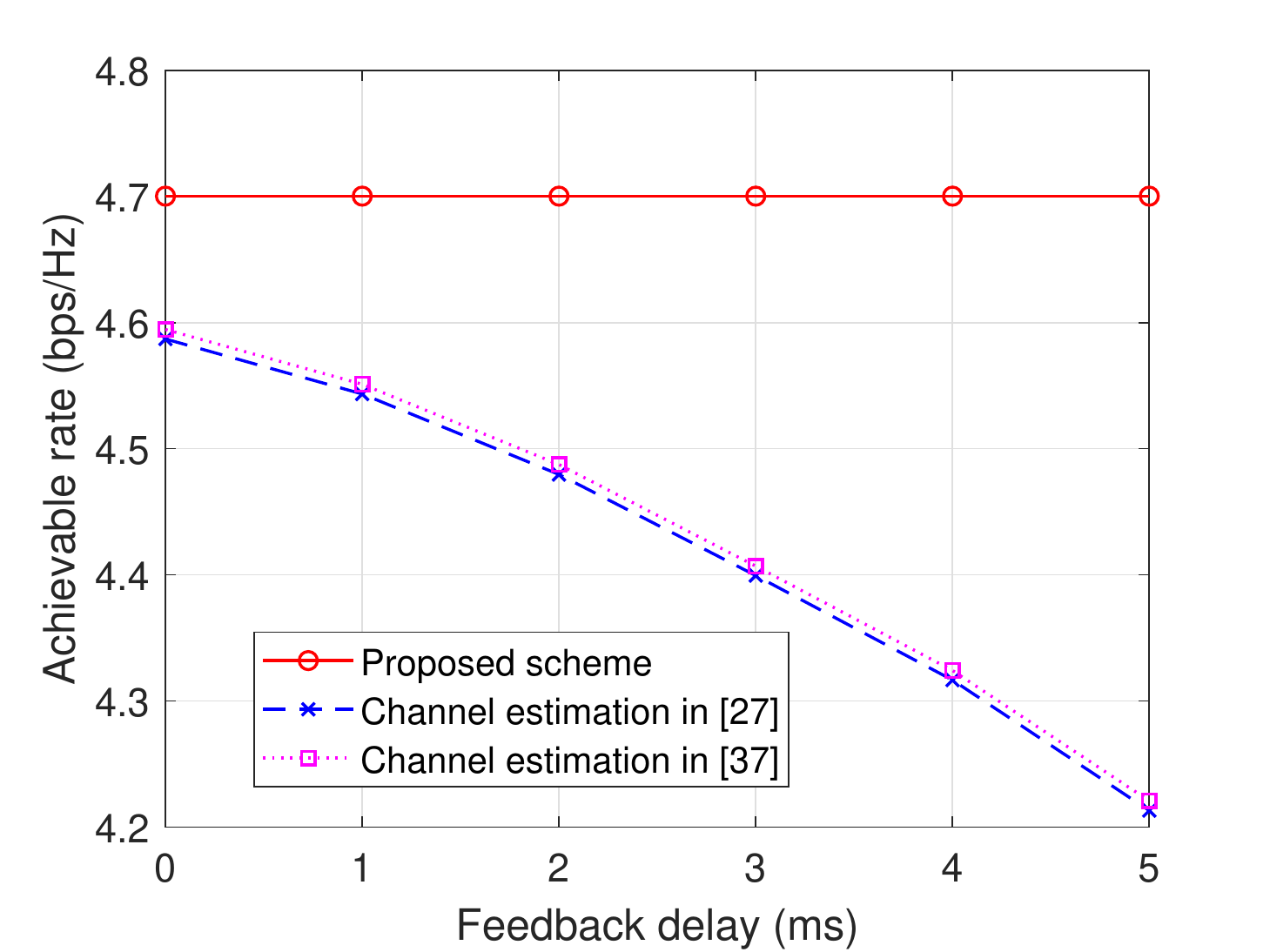}
\caption{Achievable rate versus the duration of feedback delay.}
\label{rate_fd}
\end{figure}
Finally, we show the achievable rate versus the duration of feedback delay in Fig.~\ref{rate_fd}, with $P_t = 12$~dBm, $M = 200$, $d_{\mathrm{IRS}} = 2$~m, $\tau = 12$, $N_\mathrm{B} = 16$, and $N_0 = 30$.
It is worth noting that the proposed scheme does not require real-time feedback from the BS, and thus its achievable rate is invariant to the feedback delay. In contrast, the achievable rates of the benchmark schemes in \cite{liu} and \cite{dll} both decrease with the increasing feedback delay. This is due to the fact that the cascaded BS-IRS-user channel varies rapidly over time, which renders the passive beamforming misaligned and less effective due to the outdated CSI.

\section{Conclusions}
In this paper, we investigated a new roadside IRS-aided high-mobility vehicular communication system, where a low-complexity passive beamforming design and an efficient channel estimation scheme were proposed, which require neither modification of the existing uplink transmission protocol nor real-time feedback from the BS to each serving IRS, thus making the roadside IRS-aided system practically appealing for enhancing the high-speed vehicular communication performance efficiently.
Simulation results showed that the proposed designs can effectively improve the user\textrightarrow BS channel gain distribution and thus lead to significantly enhanced communication throughput and reliability.

Although this work has demonstrated the great potential of the proposed roadside IRS-aided vehicular communication system under the simple single-user setup with frequency-flat fading channels, the results obtained can be extended to more general setups such as the multiple users, broadband system with frequency-selective fading channels, and practical IRS reflection codebook design, which are interesting as well as challenging to investigate in the future.
Moreover, more sophisticated/efficient algorithms for IRS channel estimation/prediction and passive beamforming designs are worthy of further investigation.


\end{document}